\documentclass[12pt,journal]{IEEEtran}
\usepackage{epsfig,makeidx,color}
\usepackage{graphicx}
\usepackage{amsbsy}
\usepackage{amsmath}
\usepackage{amssymb}
\usepackage{euscript}
\usepackage{ulem}

\newtheorem{mytheorem}{\bf Theorem}[section]
\newtheorem{mylemma}{\bf Lemma}

\newtheorem{myremark}{Remark}

\newtheorem{myexample}{\it Example}

\newcommand {\Define} {\stackrel {\Delta} {=}  }

\renewcommand{\baselinestretch}{1.59}
\begin{document}

%\linenumbers

\title{Improving the Performance of the Zero-Forcing Multiuser MISO Downlink Precoder through User Grouping} \author{Saif~Khan~Mohammed,~\IEEEmembership{~Member,~IEEE,}
  and~Erik G. Larsson,~\IEEEmembership{Senior~Member,~IEEE}
  \thanks{\scriptsize S. K. Mohammed is with the Department of Electrical Engineering,
Indian Institute of Technology (I.I.T.) Delhi, India and is also associated
with the Bharti school of Telecommunication Technology and Management (BSTTM), I.I.T., Delhi.
E. G. Larsson is with the
    Communication Systems Division, Dept.  of Electrical Engineering
    (ISY), Link{\"o}ping University, 581 83 Link\"oping,
    Sweden. E-mail: $\tt saifkmohammed$@$\tt gmail.com$ and $\tt
    erik.larsson$@$\tt isy.liu.se$.
    The work of Saif Khan Mohammed was partly supported by the EMR funding from the Science and
    Engineering Research Board (SERB), Department of Science and Technology (DST), Government of India. The work of E. G. Larsson was supported by the Swedish research council  and ELLIIT. This paper is a substantial extension to our conference paper presented at IEEE Globecom 2011 \cite{Gcom11}.
} }

\onecolumn
\maketitle

\vspace{-10mm}

\begin{abstract}
We consider the Multiple Input Single Output (MISO)
Gaussian Broadcast channel with $N_t$ antennas at the base station (BS) and $N_u$ single-antenna users in the downlink. We propose a novel user
grouping precoder which improves the sum rate performance of the Zero-Forcing (ZF)
precoder specially when the channel is ill-conditioned.
The proposed precoder partitions all the users into small groups of equal size.
Downlink beamforming is then done in such a way that, at each user's receiver the interference from the
signal intended for users not in its group is nulled out.
Intra-group interference still remains, and is cancelled through successive interference pre-subtraction
at the BS using Dirty Paper Coding (DPC).
The proposed user grouping method is different from user selection, since it is a method
for precoding of information to the selected (scheduled) users, and not for selecting which users
are to be scheduled.
Through analysis and simulations,
the proposed user grouping based precoder is shown
to achieve significant improvement in the achievable sum rate when compared to the
ZF precoder. When users are paired (i.e., each group has two users), the complexity of the proposed precoder is $O(N_u^3) + O(N_u^2 N_t)$ which is the same as that of
the ZF precoder.
\end{abstract}


\IEEEpeerreviewmaketitle
%%%%%%%%%%%%%%%%%%%%%%%%%%%%%%%%%%%%%%%%%%%%%%%%%%%%%%%%%%%%%%%%%%%%%%%%%%%%%%%%%%%%%%%%%%%%%%%%%
\section{Introduction}

Multiple-Input Multiple-Output (MIMO) technology holds the key to very
high throughput downlink communication in fading wireless channels by
exploiting the spatial dimension \cite{Telatar}. However most modern
MIMO wireless communication standards support a maximum achievable spectral
efficiency of around $10$ bits/sec/Hz.  This is partly due to the use of
sub-optimal orthogonal multiple access schemes like TDMA
and FDMA.
The capacity region and sum capacity
of the Gaussian MIMO broadcast channel (which models downlink
communication in modern wireless systems) is achieved by a scheme
called Dirty Paper Coding (DPC), in which all users share the same
frequency-time resource \cite{Weingarten}.  It is also known that
orthogonal access schemes (like TDMA, FDMA) are strictly sub-optimal
and achieve only a small fraction of the total sum capacity
\cite{Jindal}. However, TDMA and FDMA are still favored in practice
due to the high precoding complexity of optimal precoders like DPC.
%Also, fewer base station antennas are favored in practice, due to the
%prohibitive hardware and information precoding complexity of base stations with large number of antennas.
Apart from DPC, other near-optimal precoders like those based on vector perturbation and lattice reduction
\cite{Peel,Windpassinger} also have prohibitive complexity.
On the other hand low complexity precoders,
like ZF \cite{Baier}, MMSE
are known to
achieve poor sum rate performance especially in ill-conditioned channels.

%One of the most simple and low complexity linear precoder is the ZF precoder.
To keep the low-complexity benefit of the ZF precoder
and yet improve the overall sum rate (specially when the channel is ill-conditioned),
we propose a user grouping based precoder.
In the proposed precoder, the users are divided into {\it small} groups of equal size.
Downlink beamforming is done in such a way that, at each receiver the interference from the
signal intended for users not in its group is nulled out.
However, there still remains interference from the signal of users in the same group.
This interference is pre-cancelled at the transmitter, by performing
dirty paper coding among the users in the same group. With small groups (e.g., having only two users), dirty paper coding within each group
is {\it practically} feasible \cite{WeiYu,Sun,Shilpa}.
%Here we also make a note that, in recent literature many practical DPC schemes have been proposed
%for broadcast channels with few users, using traditional coding schemes for the
%AWGN channel.
%It is to be noted that the proposed user grouping method is not the same as user selection since
%we only consider scenarios where the users must be perpetually scheduled (e.g., delay-limited applications). In contrast, in user selection schemes \cite{TuBlum, Fuchs} only a subset of users is scheduled at any time.
Note that the proposed user grouping method is fundamentally different from user selection. User selection schemes
select a subset of users to be scheduled \cite{SHuang, Ouldooz, ZShen, TuBlum}. The base station (BS) then precodes information
only to these selected users. The proposed user grouping precoder is a method
for precoding of information to the selected users, and not for selecting which users
are to be scheduled.\footnote{\footnotesize{This distinction is the same as that between the work in \cite{Spencer} and that in \cite{ZShen}. In \cite{Spencer} the authors propose a block diagonalization method for precoding of information to already selected users, whereas in \cite{ZShen} the authors propose a method to find the subset of users to be scheduled
so that the information sum rate (using a block diagonalization precoder) is maximized.}} Note that the user grouping precoder proposed by us in this paper could be used to significantly improve the overall information sum rate performance achieved
by user selection methods which assume a ZF precoder at the BS (for example the user selection method proposed in \cite{SHuang}).  

Inter-group interference pre-cancellation
for a group of users is achieved by choosing their beamforming vectors to lie in a space orthogonal to the
space spanned by the channel vectors of the users in the other groups.
One novel aspect of the proposed precoder is that we choose the beamforming vectors
in such a way that the effective channel matrix for each group is lower triangular, which enables
successive interference pre-cancellation
within each group using DPC.
With a group size greater than one, the proposed precoder is analytically shown to achieve a sum rate greater
than that achieved by the ZF precoder. 
For a given grouping of users, the optimal power allocation is given by the waterfilling scheme.
Since the achievable sum rate of the proposed precoder is observed to be sensitive towards
the chosen grouping of users,
% and hence the achievable sum rate also needs to be maximized with respect
%to the grouping of users.
the information sum rate is jointly optimized w.r.t. both the per user power allocation
as well as the grouping.
This optimization problem is inherently complex, and therefore we propose a near-optimal
solution to it, which we refer to as JPAUGA (Joint Power Allocation and User Grouping Algorithm).

Through simulations, we show that in ill-conditioned channels the proposed precoder
with JPAUGA user grouping achieves
a sum rate significantly greater than that achieved by the ZF precoder. Further
for the special case of user pairing (i.e., two users in each group),
interference pre-cancellation needs to be performed for only one user in each group,
for which practical and near-optimal performance achieving (i.e., close to DPC) methods have been
reported \cite{WeiYu}.
Further, with user pairing the
complexity of the proposed precoder with JPAUGA user grouping is shown to have
a complexity of $O(N_u^3) + O(N_u^2 N_t)$ which is the same as the complexity
of the ZF precoder. A special case of the proposed precoder is when there is only one group
containing all the $N_u$ users. This special case has been proposed as the ZF-DP precoder
in \cite{Caire}. Though the ZF-DP precoder achieves better performance than the proposed
user grouping precoder with more than one group, it has a much higher complexity.

%The proposed user grouping based precoder achieves zero inter-group
%interference by beamforming information for a group of $g$ users in a
%direction orthogonal to the space spanned by the channel vectors of
%the remaining $N_u - g$ users not in the group. 
%Therefore the
%proposed precoder may appear similar to other block diagonalization
%based precoders reported in literature, for e.g., \cite{Spencer}. We however
We also clarify that, the proposed precoder is {\it entirely different} from
the block diagonalization based precoder proposed in \cite{Spencer}, which
considers a MIMO broadcast channel,
in which each user could have multiple receive antennas. Beamforming
vectors are chosen such that each user sees no interference from
the information intended for other users.  Hence, in the special case of MISO broadcast
channel (which we consider in this paper), the block diagonalization
precoder in \cite{Spencer} reduces to the ZF precoder.
In addition to this, the precoder that we propose
performs {\it beamforming in groups of users}
and not separately for each user.
%Another user pairing precoder has been
%proposed for the Gaussian MIMO broadcast channel in
%\cite{Hottinen}. However, in \cite{Hottinen}, only $2$ users
%share the same time-frequency resource, i.e., the medium access is
%orthogonal in groups of two users, which is a sub-optimal
%utilization of resources when compared
%to the proposed precoder where all users
%share the same time-frequency resource.

%We would also like to mention that, this paper is a substantial extension
%of a conference paper presented at IEEE Globecom 2011 \cite{Gcom11}.
The following {\it notations} have been used in this paper.
${\bf A}^H$ and ${\bf A}^T$ represent conjugate
transpose and transpose of the matrix ${\bf A}$ respectively.
For any complex number $z$, let ${z^*}$ and $\vert z \vert$ denote its complex conjugate
and absolute value respectively.
For a random variable $X$, let ${\mathbb E}[X]$ denote its expected value.
The complex and the real fields are denoted by ${\mathbb C}$ and ${\mathbb R}$ respectively.
Given a vector ${\bf x} = (x_1, x_2, \cdots, x_n)^T \in {\mathbb C}^n$, let $\Vert x \Vert \Define \sqrt{\sum_{k=1}^{n} \vert x_k \vert^2}$.
For any two real numbers $x, y \in {\mathbb R}$, let $\max(x,y)$ be equal to the maximum between $x$ and $y$.
Also, for any real $x$, $[ x ]^{+} \Define \max(x,0)$.
%${\mathbf I}_n$ denotes the $n \times n$ identity matrix.
%For any positive integer $n$, factorial of $n$ is defined as $n ! \Define n.(n-1).(n-2).\cdots.2.1$.
%Given two positive integers $n, k$ ($n \geq k$), ${n \choose k} \Define n.(n-1).\cdots.(n- k+ 1) / k!$ is the number of possible ways
%of choosing an unordered set of $k$ distinct objects from $n$ objects.
Let $\vert S \vert$ denote the cardinality (size) of the set $S$.
Given a square matrix ${\bf X}$, let $\vert {\bf X} \vert$ denote its determinant.
$\log(x)$ and $\log_2(x)$ denote the natural and base-2 logarithm of a positive real number $x$.
%\section*{Acknowledgment}


%\bibliographystyle{IEEEtran}

%%%%%%%%%%%%%%%%%%%%%%%%%%%%%%%%%%%%%%%%%%%%%%%%
\section{System model}\label{Sysmodel}
%%%%%%%%%%%%%%%%%%%%%%%%%%%%%%%%%%%%%%%%%%%%%%%%
Let ${\bf H} = ({\bf h}_1 \,,\, {\bf h}_2 \,,\, \cdots  \,,\,{\bf h}_{N_u} )^H$
represent the $N_u \times N_t$ channel matrix between the base station
and the $N_u$ single antenna users\footnote{\footnotesize {Throughout the paper, ${\bf H}$ is assumed to be full rank.}} ($N_t \geq N_u$).
The channel vector from the BS to the $k$-th user is denoted
by ${\bf h}_k^{H} \in {\mathbb C}^{1 \times N_t}$, with its $i$-th entry ${ {{h}^*_{k,i}}}$
representing the channel gain from the $i$-th transmit antenna to
the receive antenna of the $k$-th user\footnote{\footnotesize{Subsequently we shall
also refer to the receiver at the $k$-th user as the $k$-th receiver.}}.
%Let ${\bf u} = (u_1, u_2, \cdots, u_{N_u})^T \in {\mathbb C}^{N_u \times 1}$ denote the information
%symbol vector, with $u_k \in {\mathbb C}$ being the information symbol
%for the $k$-th user.
%The information symbols are assumed to be i.i.d. Gaussian distributed with mean 0 and variance 1.
The BS is assumed to have perfect channel state information (CSI).
Let ${\bf x} = (x_1, x_2, \cdots, x_{N_t})^T \in {\mathbb C}^{N_t \times 1}$
represent the transmitted vector.
The vector of received symbols ${\bf y} = (y_1, y_2, \cdots, y_{N_u})^T \in {\mathbb C}^{N_u \times 1} $
(with $y_k$ denoting the signal received by the $k$-th user) is then given by
\begin{equation}
\label{sys_model_eq}
{\bf y} = {\bf H} {\bf x} + {\bf n}
\end{equation}
where ${\bf n} = (n_1, n_2, \cdots n_{N_u})^T \in {\mathbb C}^{N_u \times 1}$ is the additive noise
vector with $n_k$ representing the noise at the $k$-th receiver. Further, each entry of ${\bf n}$ is an
i.i.d.\ ${\mathbb C}{\mathbb N}(0,1)$ random variable.
Also, the BS is subject to an average transmit power constraint
given by
\begin{equation}
\label{tx_pow_constr}
{\mathbb E}[\Vert {\bf x} \Vert^2] = P_T.
\end{equation}
%where ${\mathbb E}[.]$ denote the expectation operator, and
%$\Vert {\bf x} \Vert$ represents the Euclidean length (norm)
%of the vector ${\bf x}$.
%The rate achieved by the $k$-th user is then given by
%\begin{equation}
%R_k = I(x_k, y_k) = h(y_k) - h(y_k | x_k)
%\end{equation}
%where $h(.)$ is the differential entropy function of random variables
%with infinite support set.
%The sum rate is then given by
%\begin{equation}
%R = \sum_{k=1}^{N_u} R_k.
%\end{equation}
Due to unit variance noise, we will refer to $P_T$ as the transmit signal to receiver noise ratio (i.e., transmit SNR).
Subsequently we shall refer to the $k$-th user by ${\mathcal U}_k$.
In the proposed precoding scheme, the total set of users
${\mathcal S} = \{ {\mathcal U}_1, {\mathcal U}_2, \cdots {\mathcal U}_{N_u} \}$
is partitioned into $N_g = N_u/g$ disjoint groups of size $g$.
Let the $i$-th group of users be denoted by the ordered set
${\mathcal S}_i = \{ {\mathcal U}_{i_1}, {\mathcal U}_{i_2}, \cdots, {\mathcal U}_{i_g} \}$.
Therefore, ${\mathcal S} = \cup_{i=1}^{N_g} {\mathcal S}_i$, and ${\mathcal S}_i \cap {\mathcal S}_j = \emptyset,
\forall i \ne j$, where $\emptyset$ denotes the null set.
Also, let any arbitrary grouping of users be denoted by the unordered set ${\mathcal P} = {\Big \{} {\mathcal S}_1, {\mathcal S}_2, \cdots, {\mathcal S}_{N_g} {\Big \}}$.
For example, with $N_u = 4$ and $g=2$, one possible grouping of users is given by ${\mathcal P} = {\Big \{} \{ {\mathcal U}_1 , {\mathcal U}_4 \} , \{ {\mathcal U}_2 , {\mathcal U}_3 \} {\Big \}}$.

For notational purposes, let us denote the set of all possible
groupings of a set of $N_u$ users into groups of size $g$, by
${\mathcal A}_{N_u}^{(g)}$. For example with $N_u=4$ users and $g=2$
{\footnotesize 
\begin{eqnarray*}
{\mathcal A}_4^{(2)} & = & {\Bigg \{}
{\Big \{} \{{\mathcal U}_1,{\mathcal U}_2\}, \{{\mathcal U}_3,{\mathcal U}_4\} {\Big \}},  {\Big \{} \{{\mathcal U}_2,{\mathcal U}_1\}, \{{\mathcal U}_3,{\mathcal U}_4\} {\Big \}},{\Big \{} \{{\mathcal U}_1,{\mathcal U}_2\}, \{{\mathcal U}_4,{\mathcal U}_3\} {\Big \}}, 
{\Big \{} \{{\mathcal U}_2,{\mathcal U}_1\}, \{{\mathcal U}_4,{\mathcal U}_3\} {\Big \}}, \nonumber \\
 & & {\Big \{} \{{\mathcal U}_1,{\mathcal U}_3\}, \{{\mathcal U}_2,{\mathcal U}_4\} {\big \}},{\Big \{} \{{\mathcal U}_3,{\mathcal U}_1\}, \{{\mathcal U}_2,{\mathcal U}_4\} {\big \}}, 
 {\Big \{} \{{\mathcal U}_1,{\mathcal U}_3\}, \{{\mathcal U}_4,{\mathcal U}_2\} {\big \}},
{\Big \{} \{{\mathcal U}_3,{\mathcal U}_1\}, \{{\mathcal U}_4,{\mathcal U}_2\} {\big \}}, \nonumber \\
& & {\Big \{} \{{\mathcal U}_1,{\mathcal U}_4\}, \{{\mathcal U}_3,{\mathcal U}_2\} {\Big \}}, 
{\Big \{} \{{\mathcal U}_4,{\mathcal U}_1\}, \{{\mathcal U}_3,{\mathcal U}_2\} {\Big \}},{\Big \{} \{{\mathcal U}_1,{\mathcal U}_4\}, \{{\mathcal U}_2,{\mathcal U}_3\} {\Big \}},{\Big \{} \{{\mathcal U}_4,{\mathcal U}_1\}, \{{\mathcal U}_2,{\mathcal U}_3\} {\Big \}}   {\Bigg \}}. \nonumber
\end{eqnarray*}
}
Let ${\mathbf H[i]} \in {\mathbb C}^{(N_u - g) \times N_t}$ denote the sub-matrix of ${\bf H}$
consisting of only those rows which represent the channel vector of users {\it not} in the
set ${\mathcal S}_i$, and
let ${\mathbf G}[i] \in {\mathbb C}^{g \times N_t}$ denote the
sub-matrix containing the remaining rows of ${\bf H}$.
Specifically, if ${\mathcal S}_i = \{ {\mathcal U}_{i_1}, {\mathcal U}_{i_2}, \cdots, {\mathcal U}_{i_g}  \}$ then
\begin{equation}
\label{Gi_eq}
{\mathbf G}[i] \Define ({\bf h}_{i_1}, {\bf h}_{i_2}, \cdots ,{\bf h}_{i_g})^H.
\end{equation}
Further let ${\mathcal H}_i$ represent the subspace spanned by the rows of ${\mathbf H[i]}$, and
let ${\mathcal H}^{\perp}_i$ be the subspace orthogonal to  ${\mathcal H}_i$.
The projection matrix for the subspace ${\mathcal H}^{\perp}_i$ is denoted by
\begin{equation}
\label{Pi_def}
{\mathbf P[i]} = ({\mathbf I}_{N_t} - {\mathbf H[i]}^H({\mathbf H[i]}{\mathbf H[i]}^{H})^{-1}{\mathbf H[i]}) \in {\mathbb C}^{N_t \times N_t}.
\end{equation}
Note that ${\mathbf H[i]}{\mathbf P[i]} = 0$.
Further for the user ${\mathcal U}_{i_j}$, let ${\mathcal C}_{i_j} \subset {\mathbb C}^{N_t}$
denote the space of vectors orthogonal to the space spanned by the rows of ${\mathbf H[i]}$ and the rows of the previous $(j - 1)$ users
in the $i$-th ordered group ${\mathcal S}_i$ (i.e., ${\mathbf h}_{i_1}^H , {\mathbf h}_{i_2}^H, \cdots , {\mathbf h}_{i_{j-1}}^H$).

\section{ZF precoder and the motivation for grouping users} \label{prior_art}

The ZF precoder is a low complexity linear precoder where the information for each user is beamformed in a direction which
is orthogonal to the space spanned by the channel vectors of the remaining $N_u - 1$ users,
thereby resulting in no inter-user interference.
Hence, for any given user, its effective channel gain is proportional to the Euclidean length
of the projection of its channel vector onto the space orthogonal to the space spanned by the
channel vectors of remaining users.
In case of ill-conditioned channels, since the channel vectors of all the users are ``nearly'' linearly
dependent, the effective channel gain of each user would be small, implying low achievable rates.
Therefore it makes sense to design precoders which have a complexity similar to ZF, but which can achieve
a higher sum-rate than the ZF precoder when the channel is ill-conditioned.

By grouping users into groups of size larger than one, beamforming can
be done to nullify {\it only inter-group} interference.  Further, it
is possible to perform beamforming in such a way that the effective $
g \times g$ channel matrix for each group is lower triangular.  With
small group size and a lower triangular effective channel matrix, intra-group interference can be pre-cancelled
using {\it practical successive} dirty paper coding (DPC) at the transmitter, without any
significant increase in the required transmit power (when compared to an ideal scenario where
the effective channel matrix is diagonal, i.e., no intra-group
interference). With this
precoding method, the effective channel gain for ${\mathcal U}_{i_j}$
would be the Euclidean length of the projection of ${\bf h}_{i_j}^H$
onto the space ${\mathcal C}_{i_j}$ (i.e., user ${\mathcal U}_{i_j}$
would see interference only from the information symbols of users
${\mathcal U}_{i_{(j+1)}}, \cdots, {\mathcal U}_{i_{g}}$).

On the other hand, with the ZF
precoder, the effective channel gain is the Euclidean length of the
projection of ${\bf h}_{i_j}^H$ onto the subspace orthogonal to all
the rows of ${\bf H}$ except ${\bf h}_{i_j}^H$. (We shall subsequently
denote this orthogonal subspace by ${\mathcal H}^{\perp}_{i_j}$.)  It
is noted that ${\mathcal H}^{\perp}_{i_j} \subset {\mathcal C}_{i_j}$
whenever $g > 1$.  Since the projection of a vector onto a subspace of
some space ${\mathcal G}$ is of lesser Euclidean length than its
projection onto the space ${\mathcal G}$, it follows that the
effective channel gain for ${\mathcal U}_{i_j}$ is higher with the
proposed user grouping based precoder as compared to that with the ZF
precoder. This simple observation coupled with the availability of
practical low-complexity DPC for Gaussian broadcast channels with
a small number of users, motivates the proposed user grouping based precoder which
is presented in Section \ref{proposed_prec_ref} in more detail. For a given
user grouping the sum rate is maximized by the waterfilling power allocation
across all the users (the effective channel gain of each user is considered).

The sum rate achieved by the proposed precoder is shown to be dependent on the
chosen grouping of users. This is expected, as for example with two users having ``highly''
linearly dependent channel vectors, the information rate to these two users would be higher
when they are placed in the same group.
Therefore in Section \ref{sub_sec_partition} we propose to jointly maximize the sum rate of the proposed precoder
w.r.t. the power allocation and the possible user groupings.

\section{Proposed User Grouping based Precoder} \label{proposed_prec_ref}

%(It is easily checked that ${\mathbf H[i]}{\mathbf P[i]} = {\bf 0}$.)
%Note that ${\mathbf P[i]}$ is a Hermitian matrix and is also idempotent $i.e., {\mathbf P[i]}^2 = {\mathbf P[i]}$.
%We also observe that ${\mathbf P[i]}$ is invariant to row permutations within ${\mathbf H[i]}$.
This section is organized into several subsections.
For a given user grouping ${\mathcal P}$, we beamform information symbols in such a way that
only inter-group interference is nullified. With the proposed beamforming
the original $N_u$-user Gaussian broadcast channel is transformed into $N_g$ parallel
$g$-user Gaussian broadcast channels. This is presented in Section \ref{bmf}, where we
finally show that the proposed multiuser beamforming is such that the effective channel matrix for
each group is lower triangular. Subsequently in Section \ref{dpc_coding}, using the fact that the
effective channel is lower triangular we use Dirty Paper Coding to cancel interference
between the users within a group. We also show that for a fixed user grouping, the information sum
rate is maximized by the waterfilling power allocation. In Section \ref{zf_p} we show that the ZF
precoder is a special case of the proposed precoder with $N_u$ groups, i.e., $g=1$. We also present
expressions for the sum rate achieved by the ZF precoder. Next, in Section \ref{bet_perf} we analytically show that the proposed precoder with any arbitrary grouping having $g \geq 2$ always achieves a higher
information sum rate than the ZF precoder irrespective of the channel realization ${\bf H}$ and $P_T$. Finally, in Section \ref{mot_s}
we present an example to demonstrate the higher sum rate achieved by the proposed precoder
in comparison with the ZF precoder, with random user grouping (i.e., the user grouping
is chosen independent of the CSI). Through another example we show that random user grouping
is sub-optimal, and this motivates the problem of finding the optimal user grouping which
is discussed in Section \ref{sub_sec_partition}.

\subsection{Beamforming to cancel inter-group interference}
\label{bmf}
Let ${\mathbf u[i]} \Define (u_{i_1}, u_{i_2}, \cdots, u_{i_g})^T$
be the $g \times 1$ vector of information symbols of the users in the
$i$-th group ${\mathcal S}_i$.
The information symbols are assumed to be i.i.d. Gaussian distributed with mean 0 and variance 1.
The proposed precoder maps ${\mathbf u[i]}$ onto ${\mathbf x[i]} \in {\mathbb C}^{N_t  \times 1}$
through the linear transformation
\begin{equation}
\label{precoding_i}
{\mathbf x[i]} = {\mathbf D[i]} {\mathbf u[i]}
\end{equation}
where ${\mathbf D[i]} \in {\mathbb C}^{N_t \times g}$ is the precoding matrix for the $i$-th group of users.
The vector transmitted from the BS is then given by
\begin{equation}
\label{sum_prec}
{\bf x} = \sum_{i=1}^{N_g} {\mathbf x[i]}.
\end{equation}
Note that the transmit power constraint in (\ref{tx_pow_constr}) requires that the precoding matrices
satisfy the constraint
\begin{eqnarray}
\label{D_cnstr}
\sum_{i=1}^{N_g} \, \Vert {\mathbf D[i]} \Vert_F^2 \, & = & P_T
\end{eqnarray}
where $\Vert {\bf X} \Vert_F $ denotes the Frobenius norm of the matrix ${\bf X}$.

Let ${\mathbf y[i]} \Define (y_{i_1}, y_{i_2}, \cdots, y_{i_g})^T$ be the
$g \times 1$ vector of symbols received by the users in the $i$-th group
${\mathcal S}_i$.
Using (\ref{sys_model_eq}), (\ref{precoding_i}) and
(\ref{sum_prec}), the received vector ${\mathbf y[i]}$ is given by
\begin{eqnarray}
\label{yi_sys_eq}
{\mathbf y[i]}
&=& {\mathbf G[i]} {\Big (}{\mathbf x[i]} + \sum_{k=1, k \ne i}^{N_g} {\mathbf x[k]}{\Big )} + {\mathbf n[i]} \nonumber \\
&=& {\mathbf G[i]}{\mathbf D[i]} {\mathbf u[i]} \, + \, \sum_{k=1, k \ne i}^{N_g} {\mathbf G[i]} {\mathbf D[k]} {\mathbf u[k]}  \, +  \, {\mathbf n[i]}.
\end{eqnarray}
In (\ref{yi_sys_eq}), the term $\sum_{k=1, k \ne i}^{N_g} {\mathbf G[i]} {\mathbf D[k]} {\mathbf u[k]}$
corresponds to the interference to the users in the $i$-th group due to signals transmitted by the BS for the other $(N_g - 1)$ remaining groups. This interference can be nullified by choosing the precoding matrix ${\mathbf D[k]}$ for the $k$-th group in such a way that its columns are orthogonal to the channel vectors of all the users in the other groups. One way of achieving this as well as the power constraint in (\ref{D_cnstr})
is to have
\begin{eqnarray}
\label{Dk_eqn}
{\mathbf D[k]} & = & {\mathbf Q[k]} \, {\mathbf W[k]} \,\,\,,\,\,\, k=1, \ldots, N_g
\end{eqnarray} 
where ${\mathbf Q[k]} \in {\mathbb C}^{N_t \times g}$ is the matrix whose columns form an orthonormal basis
for the subspace ${\mathcal H}^{\perp}_k$ (i.e., the subspace of vectors orthogonal to the channel vectors of
all users in the other groups except ${\mathcal S}_k$). The matrix ${\mathbf W[k]} = \mbox {diag} (\sqrt{p_{k_1}}, \sqrt{p_{k_2}}, \cdots \sqrt{p_{k_g}})$, is the diagonal power allocation matrix for the users in the $k$-th group with $p_{k_j}$ being the power allocated to the information symbol of ${\mathcal U}_{k_j}$.
Therefore by design, we have ${\mathbf G[i]} {\mathbf Q[k]} = 0$ for all $i \ne k$, since for any $i \ne k$
the rows of ${\mathbf G[i]}$ (i.e., channel vectors of users in the $i$-th group) belong to the subspace
${\mathcal H}_k$ and the columns of ${\mathbf Q[k]}$ are orthogonal to any vector in ${\mathcal H}_k$.
This then implies that ${\mathbf G[i]} {\mathbf D[k]} = 0$ for all $i \ne k$. Using this fact in (\ref{yi_sys_eq}) we get
\begin{eqnarray}
\label{yi_sys_eq1}
{\mathbf y[i]}
&=& {\mathbf B[i]} {\mathbf u[i]} \, + \, {\mathbf n[i]}
\end{eqnarray}
where
\begin{equation}
\label{heff_i}
{\mathbf B[i]} \Define {\mathbf G[i]}{\mathbf Q[i]} {\mathbf W[i]}
\end{equation}
is the $g \times g$ effective channel gain matrix for the $i$-th group of users.
From (\ref{yi_sys_eq1}) it is clear that each group of users
does not have any interference from the other groups.
Essentially the original $N_u$ user MISO broadcast channel has been decomposed into $N_g$ parallel
non-interfering $g$-user MISO broadcast subchannels.

For the $i$-th group of users an orthonormal basis for the subspace ${\mathcal H}^{\perp}_i$ (i.e., columns of ${\mathbf Q[i]}$) can be found through the QR decomposition \cite{HornJohnson} of the matrix ${\mathbf F[i]} \Define {\mathbf P[i]} {\mathbf G}[i]^H $
which is given by
\begin{equation}
\label{qrd}
{\mathbf F[i]} = {\mathbf Q[i]} {\mathbf R[i]}.
\end{equation}
Here ${\mathbf R[i]} \in {\mathbb C}^{g \times g}$ is an upper triangular matrix with
positive diagonal entries (since ${\mathbf F[i]}$ is full rank), and ${\mathbf Q[i]} \in {\mathbb C}^{N_t \times g}$
is a matrix with orthonormal columns.
The $g$ orthonormal columns of ${\mathbf Q[i]}$
form an orthonormal basis for the space ${\mathcal H}^{\perp}_i$ since ${\mathbf H[i]} {\mathbf Q[i]} {\mathbf R[i]} = {\mathbf H[i]} {\mathbf F[i]} = {\mathbf H[i]} {\mathbf P[i]} {\mathbf G[i]}^H = 0$ and therefore ${\mathbf H[i]} {\mathbf Q[i]} = 0$. 

%Also, let the power allocated to the $i$-th group of users be $p_i \geq 0$.
%Using (\ref{precoding_i}), this then implies that
%\begin{equation}
%\label{pow_constr_i}
%\Vert {\mathbf W[i]} \Vert_F^2 = p_i, \,\,\,i=1,2,\cdots, N_g
%\end{equation}
%where $\Vert {\bf A} \Vert_F$ denotes the Frobenius norm of the matrix ${\bf A}$.
%The total power constraint in (\ref{tx_pow_constr}) can then be written as
%\begin{equation}
%\label{tot_pow_constr}
%\sum_{i=1}^{N_g} p_i = P_T.
%\end{equation}
Using (\ref{Dk_eqn}) along with the fact that the columns of ${\mathbf Q[k]}$ are orthonormal, the sum power constraint in (\ref{D_cnstr}) is given by
\begin{eqnarray}
\label{sumpow_constr_i}
\sum_{i=1}^{N_g} \, \Vert {\mathbf D}_i \Vert_F^2 & = & \sum_{i=1}^{N_g} \, \Vert {\mathbf Q}_i  {\mathbf W}_i \Vert_F^2  \nonumber \\
& = & \sum_{i=1}^{N_g} \, \mbox{Tr} {\Big (} {\mathbf W}_i^H {\mathbf Q}_i^H  {\mathbf Q}_i  {\mathbf W}_i {\Big )} \, = \, \sum_{i=1}^{N_g} \, \mbox{Tr} {\Big (} {\mathbf W}_i^H {\mathbf W}_i  {\Big )}  \nonumber \\
& = & \sum_{i=1}^{N_g} \,  \sum_{j=1}^g \, p_{i_j} \, = \, P_T
\end{eqnarray}
where we have used the fact that ${\mathbf Q[i]}$ has orthonormal columns and $\mbox{Tr}(\cdot)$ denotes
the trace operation for matrices.
Subsequently, let ${\bf p} = ( p_1, p_2, \cdots, p_{N_u} )$ denote the power allocation vector,
with $p_i$ being the power allocated to ${\mathcal U}_i$.
We next show that the effective channel gain matrix ${\mathbf B[i]}$ is a lower triangular matrix
and is equal to ${\mathbf R[i]}^H {\mathbf W[i]}$.
From the definitions of ${\mathbf P[i]}$ and ${\mathbf Q[i]}$ in (\ref{Pi_def}) and (\ref{qrd}), it is clear that
${\mathbf P[i]}$ is the projection matrix for ${\mathcal H}^{\perp}_i$ which is also the space spanned by the columns of ${\mathbf Q[i]}$ and therefore
\begin{equation}
\label{prf_1}
{\mathbf P[i]}{\mathbf Q[i]} = {\mathbf Q[i]}.
\end{equation}
Since ${\mathbf F[i]} = {\mathbf Q[i]}{\mathbf R[i]} = {\mathbf P[i]}{\mathbf G[i]}^H$,
we have
\begin{eqnarray}
\label{prf_4}
{\mathbf R[i]} & { = } & {\mathbf Q[i]}^H {\Big (} {\mathbf Q[i]} {\mathbf R[i]}  {\Big )} \, {(a) \atop = } \, {\mathbf Q[i]}^H {\mathbf F[i]} \nonumber \\
& = & {\mathbf Q[i]}^H{\mathbf P[i]}{\mathbf G[i]}^H \nonumber \\
& {(b) \atop = } & {\mathbf Q[i]}^H {\mathbf P[i]}^H {\mathbf G[i]}^H \, = \, {\Big (} {\mathbf P[i]} {\mathbf Q[i]} {\Big )}^H  \, {\mathbf G[i]}^H  \nonumber \\
& {(c) \atop = } &  {\mathbf Q[i]}^H {\mathbf G[i]}^H
\end{eqnarray}
where step (a) follows from (\ref{qrd}), step (b) follows from the fact that ${\mathbf P[i]}$
is Hermitian and step (c) follows from (\ref{prf_1}).
Using (\ref{prf_4}) in (\ref{heff_i}) we see that
${\mathbf B[i]} = {\mathbf R[i]}^H {\mathbf W[i]}$, i.e.,
the effective channel is lower triangular.
Using this expression for ${\mathbf B[i]}$ in (\ref{yi_sys_eq1}) we have
\begin{equation}
\label{final_sys_eq_i}
{\mathbf y[i]} = {\mathbf R[i]}^H {\mathbf W[i]} {\mathbf u[i]} + {\mathbf n[i]}.
\end{equation}
%\section{Achievable sum rate for the proposed precoder} \label{ach_srate_proposed_prec}
From (\ref{final_sys_eq_i}), the received signal at the
$j$-th user in the $i$-th group is given by
\begin{eqnarray}
\label{final_sys_eq_ij}
y_{i_j} &=& {\mathbf R[i]}_{(j,j)} \sqrt{p_{i_j}} u_{i_j}  + {\Big (} \overbrace{\sum_{k=1}^{(j-1)} { {{\mathbf R[i]}^*_{(k,j)}}} \sqrt{p_{i_k}} u_{i_k}}^{\mbox{Interference term}} {\Big )}  + n_{i_j} \,\,,\,\,
j=1,2,\ldots, g
\end{eqnarray}
where ${\mathbf R[i]}_{(k,j)}$ denotes the entry of ${\mathbf R[i]}$ in the $k$-th row and the $j$-th column.
%Also, ${\overline z}$ denotes the complex conjugate of the complex number $z$.
Due to the lower triangular structure of the effective channel matrix for the $i$-th group,
from (\ref{final_sys_eq_ij}), we observe that the $j$-th user in the $i$-th group (i.e., ${\mathcal U}_{i_j}$)
has interference only from the symbols of the previous $(j -1)$ users in the same group (i.e., ${\mathcal U}_{i_1}, \cdots {\mathcal U}_{i_{(j-1)}}$).

\subsection{Dirty Paper Coding to cancel intra-group interference}
\label{dpc_coding}
In the proposed coding scheme, for the $i$-th group, we start with precoding information for the first user ${\mathcal U}_{i_1}$, and since it sees
no interference from any other user, we simply use an AWGN channel code with rate
\begin{equation}
\label{usr1_rate}
r_{i_1} = \log_2 {\Big (} 1 + p_{i_1}{\mathbf R[i]}_{(1,1)}^2 {\Big )}
\end{equation}
From (\ref{final_sys_eq_ij}) it is clear that the second user ${\mathcal U}_{i_2}$, has an interference term with contribution only from the first user ${\mathcal U}_{i_1}$.
Since the BS has perfect CSI and it knows the transmitted information symbol for the first user (i.e., $u_{i_1}$), it knows the interference term
for the second user, and can therefore perform known interference pre-cancellation using
the Dirty Paper Coding scheme \cite{Costa,Caire,Erez}.
In a similar manner, for the $j$-th user ${\mathcal U}_{i_j}$, the BS can perform Dirty Paper Coding
for the known interference term which has contributions only from the previously precoded $(j-1)$ users
${\Big (} {\mathcal U}_{i_1}, {\mathcal U}_{i_2} , \cdots, {\mathcal U}_{i_{(j-1)}} {\Big )}$. The rate achieved by the $j$-th user in the $i$-th group is therefore given by
\begin{equation}
\label{usrj_rate}
r_{i_j} = \log_2 {\Big (} 1 + p_{i_j} {\mathbf R[i]}_{(j,j)}^2 {\Big )} \,\,\,,\,\,\, j \, = \, 2, 3, \ldots, g.
\end{equation}
%The sum rate achieved by the $i$-th group of users
%is therefore given by
%\begin{equation}
%\label{sum_rate_i}
%r_{i} = \sum_{j=1}^{g} r_{i_j} 
%\end{equation}
For a given grouping of users ${\mathcal P} \in {\mathcal A}_{N_u}^{(g)}$,
total power constraint $P_T$, channel realization ${\bf H}$ and power allocation vector ${\bf p}$,
the sum rate achieved by the proposed precoder is therefore given by
\begin{equation}
\label{rhpt_expr}
r({\bf H}, P_T, {\mathcal P}, {\bf p})  \Define   \sum_{k=1}^{N_u/g} \sum_{j=1}^{g} r_{k_j} 
%& = & \sum_{k=1}^{N_u/g} \sum_{j=1}^{g} r_{k_j} \nonumber \\
 =   \sum_{k=1}^{N_u/g}  \sum_{j=1}^{g}  \log_2(1 + p_{k_j} {\mathbf R[k]}_{(j,j)}^2).
\end{equation}
Maximization of $r({\bf H}, P_T, {\mathcal P}, {\bf p})$ over ${\bf p}$ yields
\begin{eqnarray}
\label{sum_rate_tot}
r({\bf H}, P_T, {\mathcal P}) & \Define & \max_{ {\bf p} \, | \, \sum_{i=1}^{N_u} p_i = P_T, \,\, p_i \geq 0}  r({\bf H}, P_T, {\mathcal P}, {\bf p})
\end{eqnarray}
%From the expression for $r({\bf H}, P_T, {\mathcal P}, {\bf p})$ in (\ref{rhpt_expr}) it is easy to see that the power allocation problem
%in (\ref{sum_rate_tot}) is equivalent to that of the optimal power allocation for a set of parallel
%channels with channel gains ${\mathbf R[k]}_{(j,j)}$.
In (\ref{sum_rate_tot}), the optimal power allocation for a given grouping of users is given by the waterfilling scheme \cite{Cover}, i.e.
\begin{equation}
\label{wfill_popt}
p_{k_j} = {\Big [} \mu - \frac{1}{{\mathbf R[k]}_{(j,j)}^2} {\Big ]}^+ \,\,,\,\, k=1,2,\ldots,N_u/g\,,\,j=1,2,\ldots,g
\end{equation}
where $\mu > 0$ is such that
\begin{equation}
\label{wfill_popt1}
\sum_{k=1}^{N_u/g} \sum_{j=1}^g p_{k_j} = P_T.
\end{equation}

\subsection{The ZF precoder: A special case of the proposed precoder}
\label{zf_p}
We note that the ZF precoder is a special case of the proposed user grouping scheme with $g=1$, i.e., $N_u$
groups with one user per group.
%Since $g=1$ for the ZF precoder, the number of possible groupings
%i.e., ${\mathcal A}_{N_u}^{[g)} = 1$.
Subsequently, for $g=1$ (i.e, the ZF precoder), we shall denote the optimal waterfilling power allocation
(given by (\ref{wfill_popt}) and (\ref{wfill_popt1})) by ${\bf p}^* = (p_1^* , p_2^*, \cdots, p_{N_u}^*)$.
The sum rate achieved by the ZF precoder can be shown to be
\begin{equation}
\label{zf_prec_eq6}
C_{\mbox{ZF}}({\bf H} \,,\, P_T) = \sum_{i=1}^{N_u}  \log_2 {\Big (} 1 +  \frac {p_i^*} { [ ({\bf H}{\bf H}^{H})^{-1} ]_{(i,i)}  } {\Big )} 
\end{equation}
where ${\bf p}^*$ is given by
\begin{eqnarray}
\label{zf_prec_eq5}
p_i^{*}
& = & {\Big [}  \lambda -  [ ({\bf H}{\bf H}^{H})^{-1} ]_{(i,i)}   {\Big ]}^+ \,,\, \forall i=1,2,\ldots, N_u
\end{eqnarray}
The variable $\lambda > 0$ is chosen such that
\begin{equation}
\label{waterfill_zf}
\sum_{i=1}^{N_u} p_i^{*} = P_T.
\end{equation}
The other special case is for
$g=N_u$, i.e., only one group consisting of all the $N_u$ users. This has been discussed in detail in \cite{Caire} as the ZF-DP precoder.
%Further, the number of possible ordered groupings is $N_u !$, and therefore finding the optimal grouping would also be prohibitive for large $N_u$.

\subsection{The proposed precoder achieves a higher information rate than the ZF precoder}
\label{bet_perf} 
The following theorem shows that
irrespective of the channel realization ${\bf H}$ and $P_T$, the sum rate achieved by the proposed precoder
with any arbitrary user grouping having $g \geq 2$ is greater than that achieved by the ZF precoder (i.e., proposed precoder with $g=1$).
\begin{mytheorem}\label{th1}
Let ${\mathcal P} \in {\mathcal A}_{N_u}^{g}$ be any arbitrary user grouping with $g \geq 2$.
Then
\begin{eqnarray}
\label{sum_cap_g2_anypair}
r({\bf H}, P_T, {\mathcal P}) & \geq & C_{\mbox{ZF}}({\bf H} \,,\, P_T) 
%r({\bf H}, P_T, {\mathcal P}, {\bf p}^*) = \sum_{k=1}^{N_u/g}  \sum_{j=1}^{g}  \log_2(1 + p_{k_j}^{*} {\mathbf R[k]}_{(j,j)}^2) \geq  C_{\mbox{ZF}}.
\end{eqnarray}
holds for any channel realization ${\bf H}$ and $P_T$.
\end{mytheorem}

{\it Proof} --
See Appendix~\ref{th1_proof}.
$\hfill\blacksquare$

In this following we illustrate the effectiveness of the proposed idea of grouping users through an example where for a Rayleigh fading channel we show
that for any $P_T$ the ergodic sum rate (i.e, sum rate averaged over all realizations of $\bf H$) achieved by the proposed precoder (with $g=2$ and random user grouping) is always
greater than that achieved by the ZF precoder.
We will also show that to achieve a given fixed sum rate, the ZF precoder asymptotically (i.e., as $P_T \rightarrow \infty$) requires about $2.17$ dB more power than the proposed precoder (with $g=2$ and random user grouping). 

\begin{myexample}\label{ex1}
Let $N_t = N_u$ and the entries of ${\bf H}$ be i.i.d. Rayleigh faded with each entry distributed as a
circular symmetric complex Gaussian random variable having zero mean and unit variance.
Let
\begin{eqnarray}
d(P_T,N_u) & \Define & {\mathbb E}_{{\bf H}} {\Big [} r({\bf H} , P_T, {\mathcal P}, {\bf p}) \, - \,  C_{\mbox{ZF}}({\bf H} \,,\, P_T)  {\Big ]}
\end{eqnarray}
denote the difference between the ergodic sum rates achieved by the ZF precoder and that achieved by the proposed precoder (with $g=2$). 
Further, for the proposed precoder, let the user pairs (since $g=2$) be formed randomly (random grouping), i.e.,
the pairing of users is assumed to be independent of the channel realization ${\bf H}$.
The power allocation vector ${\bf p}$ for the proposed precoder is assumed to be uniform, i.e., $p_i = P_T/N_u\,,\,i=1,2,\ldots,N_u$.\footnote{\footnotesize {
It is to be noted that this is justified at high SNR ($P_T \rightarrow \infty$) since the optimal waterfilling power allocation is almost the same as uniform power allocation.}}

\begin{mylemma}
\label{lemma_1}
Under the above assumptions, $d(P_T,N_u)$ can be bounded as follows
\begin{equation}
\label{diff_erg_sum_rate6}
\frac{N_u}{2} \log_2(e) {\Big (} 1 - \frac{N_u}{P_T} \log(1 + \frac{P_T}{N_u}) {\Big )}  \,  < \,  d(P_T,N_u) \,  < \, \frac{N_u}{2} \log_2(e) {\Big (} 1 - \frac{N_u}{2P_T} \log(1 + \frac{2P_T}{N_u}) {\Big )}.
\end{equation}
\end{mylemma}

{\it Proof} --
See Appendix~\ref{appen_th_analysis}. $\hfill\blacksquare$
\end{myexample}

\begin{myremark}
We firstly note that both the upper and lower bounds in (\ref{diff_erg_sum_rate6}) are strictly positive for all $P_T > 0$. This is because $g(x) \Define x \, - \, \log(1 + x)$ is strictly positive for all $x > 0$,
and the lower and upper bounds in (\ref{diff_erg_sum_rate6}) are $\frac{g(P_T/N_u)}{P_T/N_u}$ and $\frac{g(2P_T/N_u)}{2P_T/N_u}$ respectively.\footnote{\footnotesize{Note that $g(x=0) = 0$ and its first derivative $\frac{d g(x)}{d x} = \frac{x}{1 + x} \, > \, 0$ for all $x > 0$. This implies that $g(x) > 0$ for all $x > 0$.}}
For a fixed $N_t = N_u$, the lower and upper bounds in (\ref{diff_erg_sum_rate6}) can be shown to converge to $N_u \log_2(e)/2$ as $P_T \rightarrow \infty$, which implies that at sufficiently high SNR, {\it by randomly pairing users the proposed precoder can achieve an ergodic sum rate which
is $N_u \log_2(e)/2$ bits per channel use (bpcu) greater than the ergodic sum rate achieved by the ZF precoder}.
Further, at high SNR the slope of the sum rate achieved by the ZF precoder w.r.t. $\log(P_T)$ is $N_u\log_2(e)$.
This then implies that at high SNR,
the ZF precoder needs roughly $10\log_{10}(\sqrt{e}) = 2.17$ dB more power than that required by the proposed precoder with ($g=2$ , random grouping) to achieve a given ergodic sum rate.
An important observation on this result is that, the asymptotic SNR gap of $2.17$ dB is {\it independent of $N_u$}.

The above analysis shows that, even with random user grouping, the proposed grouping based precoder is more power efficient than the ZF precoder.
$\hfill\square$
\end{myremark}

\subsection{Motivating the need for ``optimal'' user grouping}
\label{mot_s}

So far we have not bothered much about the choice of user grouping.
The following example shows the sensitivity
of the proposed precoder w.r.t. the chosen user grouping.
This then motivates us to choose the user grouping which maximizes the sum rate.

\begin{myexample}
\label{ex2}
In this example we consider a $N_t = N_u = 6$ Gaussian broadcast channel whose
channel matrix is ill-conditioned and is given by
{\footnotesize
\begin{equation}
\label{illcond_eq}
{\bf H}_{ex} = 
\left [ 
\begin{array}{cccccc}
\frac{1}{2} & 0 & 0 & -\frac{1}{2} & \frac{1}{\sqrt{2}} & 0 \\
0  & \frac{1}{2}   &  -\frac{1}{\sqrt{2}}  & \frac{1}{2}  & 0 & 0 \\
0 & -\frac{1}{2}  & 0 & 0 & -\frac{1}{\sqrt{2}} &  \frac{1}{2} \\
-\frac{1}{2} & 0 & 0 & \frac{1}{\sqrt{2}} & 0 &  -\frac{1}{2} \\
\frac{1}{2} & 0 & \frac{1}{\sqrt{2}}  & 0 & \frac{1}{{2}} & 0 \\
0 & 0 & 0 & -\frac{1}{\sqrt{2}} & -\frac{1}{{2}} & \frac{1}{{2}}
\end{array} \right ].
\end{equation}
\normalsize}
The ordered singular values of ${\bf H}_{ex}$ are $(1.56, 1.48, 0.97, 0.54, 0.38, 0.028)$. 
In Fig.~\ref{fig_0}, we plot the sum rate $r({\bf H},P_T,{\mathcal P})$ as a function of all the possible groupings ${\mathcal P} \in {\mathcal A}_6^2$ (i.e., with $g=2$)
for a fixed ${\bf H} = {\bf H}_{ex}$ and $P_T=10$ dB.
%Note that $r({\bf H},P_T,{\mathcal P})$ is defined in (\ref{sum_rate_tot}).
For a given grouping of users, power allocation is given by the optimal waterfilling scheme in (\ref{wfill_popt}) and (\ref{wfill_popt1}).
As observed in Fig.~\ref{fig_0}, large variations in the achievable sum rate suggests its sensitivity towards the chosen grouping of users.$\hfill\square$
\begin{figure}[t]
\begin{center}
\hspace{-1mm}
\epsfig{file=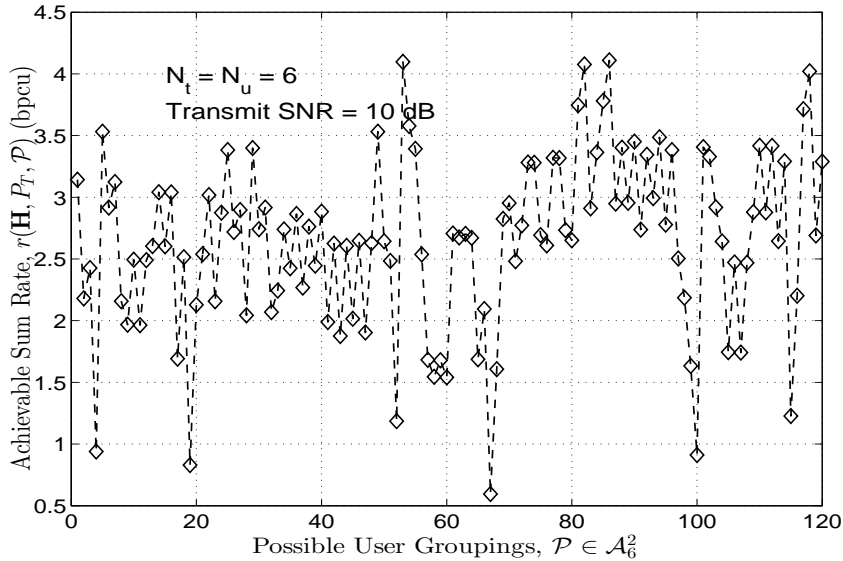, width=120mm,height=80mm}
\end{center}
\vspace{-3mm}
\caption{Sensitivity of the achievable sum rate towards the chosen grouping of users. $N_t=N_u=6$  and $g=2$. MISO broadcast channel given by (\ref{illcond_eq}). The number of possible groupings is $\vert {\mathcal A}_6^2\vert = 120$.}
\label{fig_0}
\vspace{-5mm}
\end{figure}
\end{myexample}

\begin{figure}[t]
\begin{center}
\hspace{-1mm}
\epsfig{file=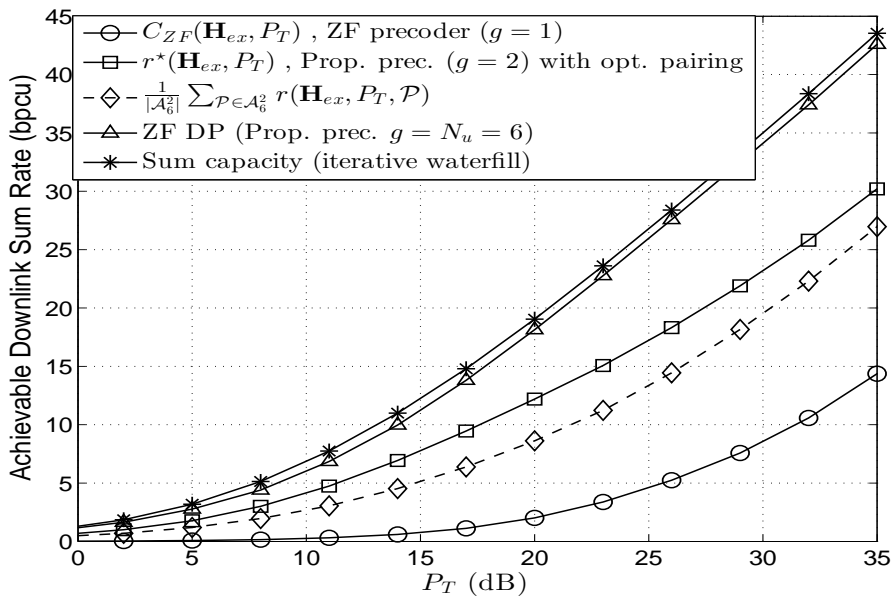, width=120mm,height=80mm}
\end{center}
\vspace{-3mm}
\caption{Comparison of the sum rates achieved by the proposed user grouping precoder and the ZF precoder for the broadcast channel in (\ref{illcond_eq}).}
\label{fig_m2}
\vspace{-5mm}
\end{figure}

Motivated by the sensitivity of the proposed precoder w.r.t. user grouping we define the optimal user
grouping as one which maximizes the sum rate. The optimal user grouping is clearly a function of
$({\bf H}, P_T)$ and is given by
\begin{eqnarray}
\label{opt_grp}
{\mathcal P}^{{ \bigstar}}({\bf H}, P_T) & \Define & \arg \max_{{\mathcal P} \in {\mathcal A}^g_{N_u}} \, r({\bf H}, P_T, {\mathcal P})
\end{eqnarray}
where $r({\bf H}, P_T, {\mathcal P})$ is given by (\ref{sum_rate_tot}).
The corresponding optimal sum rate of the proposed precoder is denoted by
\begin{eqnarray}
\label{opt_grp_sum_rate}
r^{\bigstar}{\Big (}{\bf H}, P_T {\Big )}  \Define  r{\Big (}{\bf H}, P_T, {\mathcal P}^{\bigstar}({\bf H}, P_T){\Big )}.
\end{eqnarray}
For the $6 \times 6$ channel in (\ref{illcond_eq}), we numerically compute the optimal user grouping for the
proposed precoder with $g=2$ and compare the resulting optimal sum rate with the sum rate achieved by the
ZF precoder i.e., $C_{\mbox{ZF}}({\bf H}_{ex}, P_T)$. This comparison is depicted graphically as a function
of $P_T$ in Fig.~\ref{fig_m2}. We also plot the information sum rate of the proposed precoder averaged over all
possible groupings (see the curve marked with diamonds). It is observed that indeed optimal user grouping results in significant improvement in sum rate. As an example, at $P_T = 10$ dB the information sum rate of the ZF precoder is only $0.31$ bpcu when compared to $4.75$ bpcu achieved by the proposed precoder with optimal user grouping. Also with random user grouping (curve marked with diamonds) the average information sum rate achieved by the proposed precoder is $3$ bpcu at $P_T = 10$ dB. Therefore, in ill-conditioned channels it appears that choosing the optimal grouping
can lead to significant improvement in the sum rate performance of the proposed precoder.
Note that the sum rate of the proposed user grouping scheme is significantly
higher than that of the ZF precoder even for small $g=2$. Exhaustive simulations have revealed that the sum rate of the proposed user grouping scheme increases with increasing $g$.  

In Fig.~\ref{fig_m2} we also plot the sum capacity\footnote{\footnotesize{The sum capacity of the broadcast channel is computed using the sum power iterative waterfilling method proposed in \cite{Nihar}.}} of the multiuser channel in (\ref{illcond_eq}) and the sum rate achieved by the ZF-DP precoding scheme (i.e., special case of the proposed user grouping scheme with $g= N_u = 6$). We observe that the ZF-DP scheme is near sum capacity achieving and has a better sum rate performance than the proposed user grouping precoder with $g=2$ (optimal pairing). However, the ZF-DP precoder achieves this better performance at the cost of a significantly higher complexity and other disadvantages when compared to the proposed user grouping precoder with $g=2$, as is discussed in the following.

In ZF-DP (i.e., proposed user grouping precoder with $g=N_u$) successive DPC has to be performed for $(N_u - 1)$ users, whereas
when $g=2$ successive DPC needs to be performed for only $N_u/2$ users (only for the second user in each group).
With successive DPC, the power of the known interference signal due to other users will increase with the user index, i.e., the first user to be precoded will not see any interference, the second user will see interference only from the first user, the
third user will see interference from both the first and the second user, and so on \cite{Caire}.
With $g=2$, DPC is performed only for the second user in each group, and therefore
the interference power is roughly of the same order as the power of the useful information symbol. On the other hand for ZF-DP ($g=N_u$),
the last user to be precoded needs to perform DPC for interference from all the previous $(N_u-1)$ users.
Hence the interference power for each successive DPC is expected to be higher for the ZF-DP precoder in comparison to the proposed precoder with $g=2$. This larger interference
power will lead to increase in complexity of known practical near-optimal-DPC schemes.
As an example, in \cite{WeiYu} it is mentioned that with increasing interference power the
size of the channel code alphabet set (constellation) has to be increased in order to ensure that the interference signal lies entirely inside
the expanded constellation. This expansion in the constellation will also increase the dynamic range of the received signal at the user end, which can then increase the design complexity of the receiver. In general it is expected that increasing $g$ will increase the
sum rate performance of the proposed precoder, but at the cost of higher complexity.

\section{Partitioning users into groups} \label{sub_sec_partition}

For small $N_u$, (\ref{opt_grp}) can be solved simply by brute-force enumeration of all possible groupings.
However, for large $N_u$, the combinatorial nature of the problem makes it inherently complex
to solve by brute-force enumeration.\footnote{\footnotesize{The number of possible groupings, i.e.,
$\vert {\mathcal A}_{N_u}^g \vert = N_u ! / {\Big (} (N_u / g) ! {\Big )}$ grows exponentially with $N_u$ for a fixed $g$. For example with $g=2$ and even $N_u$, $\vert {\mathcal A}_{N_u}^g \vert \, = \, 2^{N_u/2} \, (N_u - 1) \cdot (N_u -3) \cdots 3 \cdot 1$.}}
Therefore for large $N_u$ we propose an iterative ``Joint Power Allocation and User Grouping Algorithm'' (JPAUGA), which solves (\ref{opt_grp}) approximately.
Numerical results demonstrate that JPAUGA achieves an information rate close
to the optimal $r^{\bigstar}({\bf H}, P_T)$.

Let ${\mathcal P}^{(q)}$ be the user grouping after the $q$-th iteration of JPAUGA.
Similarly, let ${\bf p}^{(q)}$ be the power allocation after the $q$-th iteration of JPAUGA.
JPAUGA starts with initializing the power allocation to be the ZF power allocation i.e.,
${\bf p}^{(0)} = {\bf p}^{*}$ (see Section \ref{zf_p}). In the $q$-th iteration ($q=1,2,\ldots, \max_{itr}$), we firstly find
the user grouping ${\mathcal P}^{(q)}$ which approximately maximizes the information sum rate with
power allocation fixed to its values at the end of the $(q-1)$-th iteration, i.e.,
${\bf p} = {\bf p}^{(q-1)}$. That is, ${\mathcal P}^{(q)}$ is an approximate solution
to the problem
\begin{eqnarray}
\label{max_grp}
\arg \max_{{\mathcal P} \in {\mathcal A}_{N_u}^g} \, r{\Big (} {\bf H}, P_T, {\mathcal P}, {\bf p}^{(q-1)}  {\Big )}
\end{eqnarray}
In Section \ref{guga_ref} we propose an approximate solution to (\ref{max_grp}), called ``Generalized User Grouping Algorithm'' (GUGA).
After computing ${\mathcal P}^{(q)}$ using GUGA, the power allocation for the
$q$-th iteration, i.e., ${\bf p}^{(q)}$ is given by the waterfilling scheme with user grouping fixed
to ${\mathcal P}^{(q)}$ (see (\ref{wfill_popt}) and (\ref{wfill_popt1})).
The proposed iterative algorithm JPAUGA then moves to the $(q+1)$-th iteration.

Due to alternating maximization of the information sum rate w.r.t. user grouping
and power allocation, it is clear that the information sum rate increases
successively from one iteration to the next,
i.e., $r{\Big (}{\bf H}, P_T, {\mathcal P}^{(q+1)}, {\bf p}^{(q+1)}{\Big )} \geq  r{\Big (}{\bf H}, P_T, {\mathcal P}^{(q)}, {\bf p}^{(q)}{\Big )}$.
The algorithm terminates either after a fixed number of iterations (e.g., $\max_{itr}$)
or till the relative iteration-by-iteration improvement in the information sum rate i.e., ${\Big [} r{\Big (}{\bf H}, P_T, {\mathcal P}^{(q+1)}, {\bf p}^{(q+1)}{\Big )} \, - \,  r{\Big (}{\bf H}, P_T, {\mathcal P}^{(q)}, {\bf p}^{(q)}{\Big )} {\Big ]} /r{\Big (}{\bf H}, P_T, {\mathcal P}^{(q)}, {\bf p}^{(q)}{\Big )}$ falls below a certain
pre-determined threshold.

\subsection{\bf Generalized User Grouping Algorithm - GUGA } \label{guga_ref}

In this section we discuss the problem
of finding the user grouping which maximizes the information sum rate
for a fixed $({\bf H}, P_T, {\bf p})$, i.e.,
\begin{equation}
\label{max_grp1}
 \arg \max_{{\mathcal P} \in {\mathcal A}_{N_u}^g} \, r{\Big (} {\bf H}, P_T, {\mathcal P}, {\bf p} {\Big )}.
\end{equation}
This problem is combinatorial in nature and it appears that
finding the optimal user grouping would be prohibitive for large $N_u$.
Therefore in the following we propose a low complexity approximate solution to (\ref{max_grp1}), called
``GUGA''.
 
Before discussing GUGA in detail, for any arbitrary user grouping ${\mathcal P} = \{  {\mathcal S}_1 , \cdots, {\mathcal S}_{N_g} \}$ we define
the rate of the $k$-th group of $g$ users i.e.,
${\mathcal S}_k = \{ {\mathcal U}_{k_1}, {\mathcal U}_{k_2}, \cdots , {\mathcal U}_{k_g}  \}$
by\footnote{\footnotesize {We remind the reader that ${\mathbf R[k]}$ is implicitly dependent on the chosen grouping.}}
\begin{equation}
\label{weight_func_g}
{\mathcal I}({\mathcal S}_k)  \Define  \sum_{j=1}^{g} \log_2 (1 + p_{k_j} {\mathbf R[k]}_{(j,j)}^2).
\end{equation}
The optimization problem in (\ref{max_grp1}) can therefore be expressed as
\begin{equation}
\label{opt_weight_func_g}
 \arg \max_{{\mathcal P} = \{ {\mathcal S}_1 \,,\, {\mathcal S}_2 \,,\, \cdots \,,\, {\mathcal S}_{N_g}  \}  \in {\mathcal A}_{N_u}^{g}}  \sum_{k=1}^{N_u/g}  {\mathcal I}({\mathcal S}_k).
\end{equation}

%{\underline {\bf An iterative near-optimal greedy algorithm for solving (\ref{opt_weight_func_g})}}
The proposed GUGA algorithm is an iterative greedy algorithm. Let the set of active users after the $k$-th iteration
be denoted by ${\mathbb V}^{(k)} \subset {\mathcal S}$.
In the $(k+1)$-th iteration, a subset of ${\mathbb V}^{(k)}$ containing $g$ users is chosen to be the
$(k+1)$-th group of users.
Let ${\mathbb E}^{(k)}$ denote the set of all possible {\it ordered} subsets of ${\mathbb V}^{(k)}$ of size $g$.
That is
\begin{equation}
\label{ek_def_g}
{\mathbb E}^{(k)} \Define {\Big \{} s \subset {\mathbb V}^{(k)} \, | \, \vert s \vert = g {\Big \}}.
\end{equation}
%The cardinality of ${\mathbb E}^{(k)}$ is therefore given by
%\begin{equation}
%\label{cardinal_ek}
%\vert {\mathbb E}^{(k)} \vert = g ! {{\vert {\mathbb V}^{(k)} \vert} \choose  g}   
%\end{equation}
Starting with the $k$=$0$-th iteration the set ${\mathbb V}^{(0)} = {\mathcal S}$ (i.e., all users are active) and ${\mathbb E}^{(0)}$ is the set of all possible ordered subsets of ${\mathcal S}$ of size $g$.
In the $(k+1)$-th iteration, the proposed algorithm finds the group of $g$-users in ${\mathbb E}^{(k)}$ having the maximum rate. This group is then chosen to be the $(k+1)$-th group of users i.e.
\begin{equation}
\label{max_edge_g}
{\Tilde {\mathcal S}}_{k+1} = \{ {\mathcal U}_{{(k+1)}_1}, {\mathcal U}_{{(k+1)}_2}, \cdots, {\mathcal U}_{{(k+1)}_g} \} \Define \arg \max_{s \in {\mathbb E}^{(k)} } {\mathcal I}(s)
\end{equation}
where ${\mathcal I}(.)$ is given by (\ref{weight_func_g}).
Let ${\mathcal T}^{(k+1)}  \subset  {\mathbb E}^{(k)}$ be the set of groups of size $g$ having at least one user
in the set ${\Tilde {\mathcal S}}_{k+1}$. That is
\begin{equation}
\label{tkp1_g}
{\mathcal T}^{(k+1)} \Define
{\Big \{} s \,\,|\,\, s \in {\mathbb E}^{(k)} \,\,\mbox{and}\,\, {\mathcal U}_{{(k+1)}_j} \in  s \,\,\mbox{for some} \,\, j {\Big \}} 
\end{equation}
where ${\mathcal U}_{{(k+1)}_j}$ is the $j$-th user in the ordered set ${\Tilde {\mathcal S}}_{k+1}$.
After the $(k+1)$-th iteration, the users ${\mathcal U}_{{(k+1)}_j}, j=1,2,\ldots,g$ are removed from the active set of users, i.e.
\begin{equation}
\label{modif_V_g}
{\mathbb V}^{(k+1)}  =  {\mathbb V}^{(k)} \, \setminus \,  {\Tilde {\mathcal S}}_{k+1}
\end{equation}
where `` $\setminus$ '' denotes the minus/difference operator for sets. 
From (\ref{modif_V_g}) and the definition of ${\mathbb E}^{(k)}$ in (\ref{ek_def_g}) we therefore have
\begin{equation}
\label{modif_E_g}
{\mathbb E}^{(k+1)}  =  {\mathbb E}^{(k)} \, \setminus \,  {\mathcal T}^{(k+1)}.
\end{equation}
The algorithm then moves on to the $(k+2)$-th iteration.
Since there are totally $N_u$ users and therefore $N_u/g$ groups, it is evident that
the algorithm terminates after the $N_g = (N_u/g)$-th iteration.
The proposed grouping of users is then given by
\begin{equation}
\label{proposed_pairing_g}
{\Tilde {\mathcal P}} = \{  {\Tilde {\mathcal S}}_1 \,,\, {\Tilde {\mathcal S}}_2   \,,\, \cdots \,,\,  {\Tilde {\mathcal S}}_{N_g}  \}
\end{equation}
For the sake of clarity, in Appendix \ref{app_num_ex} we present a numerical example to
illustrate GUGA.

{{\bf Complexity of GUGA}}

%\label{cmplx_guga}
%We firstly discuss the complexity of the proposed user grouping algorithm and the corresponding
%precoder.  
%From (\ref{sum_prec}) and (\ref{precoding_i}), it is clear that
%for the $k$-th group of users, the orthogonal beamforming matrix
%${\mathbf Q[k]}$ has to be computed every time the channel changes. 
%In Appendix \ref{th3}, we discuss a low complexity technique to compute ${\mathbf Q[k]}$
%and also the upper triangular matrix ${\mathbf R[k]}$ for any arbitrary grouping of users.
The proposed user grouping algorithm (GUGA) needs to initially compute the rate of all possible subsets of ${\mathcal S}$ of size $g$.
For a given group, its rate is a function of the corresponding upper triangular matrix representing the effective channel for that group.
In Appendix \ref{subsec_Rk}, it is shown that starting with $({\bf H}{\bf H}^H)^{-1}$, the complexity of computing the effective upper triangular matrix for a given group is $O(g^3)$. From (\ref{weight_func_g}) it then follows that for a given power allocation, computing the rate ${\mathcal I}({\mathcal S}_k)$ for any arbitrary group of users ${\mathcal S}_k$ has a complexity of $O(g^3)$. Since there are $O({N_u}^g)$ possible ordered groups/subsets of ${\mathcal S}$ of size $g$ (i.e.,  $\vert {\mathbb E}^{(0)} \vert = O({N_u}^g)$), the complexity of computing the rate of all possible
groups/subsets of ${\mathcal S}$ is $O(g^3 {N_u}^g)$.
In the $(k+1)$-th iteration of GUGA, we then find the group of users having the maximum rate
among all possible groups in ${\mathbb E}^{(k)}$ (see (\ref{max_edge_g})).
The complexity of $N_g = N_u/g$ iterations of GUGA is therefore $O(N_u^{g+1})$.
Hence we can conclude that the total complexity of GUGA is $O(g^3 N_u^g) + O(N_u^{g+1})$.  

%Since there are totally $O(g! N_u^g)$ possible groups, computing the weights of all possible groups of size $g$ has a total complexity of $O(g! g^3 N_u^g) + O(g! g N_u^{(g+1)})$.
%In each iteration, the proposed user grouping algorithm
%finds the group with the maximum weight from the set of active groups (see (\ref{max_edge_g})).
%To accomplish this, a simple linear search over the set of groups would result in a complexity
%of $O(g! N_u^g)$. Since there are exactly $N_u/g$ iterations,

\subsection{Complexity of the proposed precoder based on JPAUGA}
\label{cmplx_jpauga}
The whole precoding operation can be broadly divided into two phases.
In the first phase, JPAUGA is used to compute the user grouping
and the power allocation between users. Then in the second phase,
using the JPAUGA user grouping and power allocation, the information for
different groups is beamformed in orthogonal directions and information
within each group is precoded using DPC.

For the first phase, we need to firstly compute ${\Big (} {\bf H} {\bf H}^H {\Big )}^{-1}$
which has a complexity of $O(N_u^3) + O(N_u^2 N_t)$.
Through numerical simulations we have observed that JPAUGA converges
very fast, and few iterations (less than five) are required irrespective of $(N_u, N_t)$.
The complexity of computing the optimal power allocation for a given
user grouping is $O(N_u^2)$ (see (\ref{wfill_popt}) and (\ref{wfill_popt1})). 
Since each JPAUGA iteration consists of one instance of GUGA followed
by waterfilling power allocation, it follows that the total
complexity of JPAUGA
is $O(N_u^3) + O(N_u^2 N_t) + O(g^3 N_u^g) + O( N_u^{g+1})$.

For the second phase, the complexity of computing the beamforming matrix
for a single group is $O(g^3) + O(g^2 N_u) + O(g N_u N_t) + O(g^2 N_t)$ (see Appendix \ref{subsec_Fk}).
Therefore the complexity of computing the beamforming matrices for all the $N_g = N_u/g$
groups is $O(g^2 N_u) + O(g N_u^2) + O( N_u^2 N_t) + O(g N_u N_t)$.
The complexity of beamforming the information symbols onto the
transmit vector is $O(N_tN_u)$ (see (\ref{precoding_i}) and (\ref{sum_prec})).
Additionally, we would also require to perform DPC for $(g-1)$ users in each group.
Therefore, the total complexity of the second phase would be $O(g^2 N_u) + O(g N_u^2) + O( N_u^2 N_t) + O(g N_u N_t)$
plus the complexity of performing DPC for $N_g$ $g$-user MISO-broadcast channels.

The total complexity of the proposed precoder based on JPAUGA (both first and second phase) is therefore
$O(g^2 N_u) + O(g N_u^2) +  O(N_u^3) + O( N_u^2 N_t) + O(g N_u N_t) + O(g^3 N_u^g) + O( N_u^{g+1})$ plus the complexity of performing DPC for $N_g$ $g$-user MISO-broadcast channels.

\begin{myremark}
\label{remark_same_cmplx}
For small values of $g$ (e.g., $g=2$) the effective $g \times g$ lower triangular channel matrix
is small enough so that practical near-optimal (i.e., close to DPC) performance achieving schemes can be applied.
For example, with $g=2$, due to the lower triangular nature of the effective channel matrix,
the first user in each group gets its information symbol interference free, but the second user gets its
information symbol along with some interference from the first user's information symbol.
However since this interference is already known at the BS, near-optimal interference pre-subtraction can be performed at practical complexity as shown in \cite{WeiYu}.

Also with $g=2$ the complexity of the proposed JPAUGA and group-wise beamforming is
$O(N_u^3) + O(N_u^2N_t)$, which is the same as the complexity of the ZF precoder.
$\hfill\square$
%It can therefore be concluded that with $g=2$ the proposed JPAUGA based user grouping
%precoder achieves significant performance improvement compared to the ZF precoder, at similar
%complexity.

%It is also noted that the overall complexity increases only linearly with $N_t$, which therefore
%makes the proposed precoder attractive for large/massive multiuser MISO systems.  
\end{myremark}

\section{Simulation results} \label{simu}

\begin{figure}[t]
\begin{center}
\hspace{-1mm}
\epsfig{file=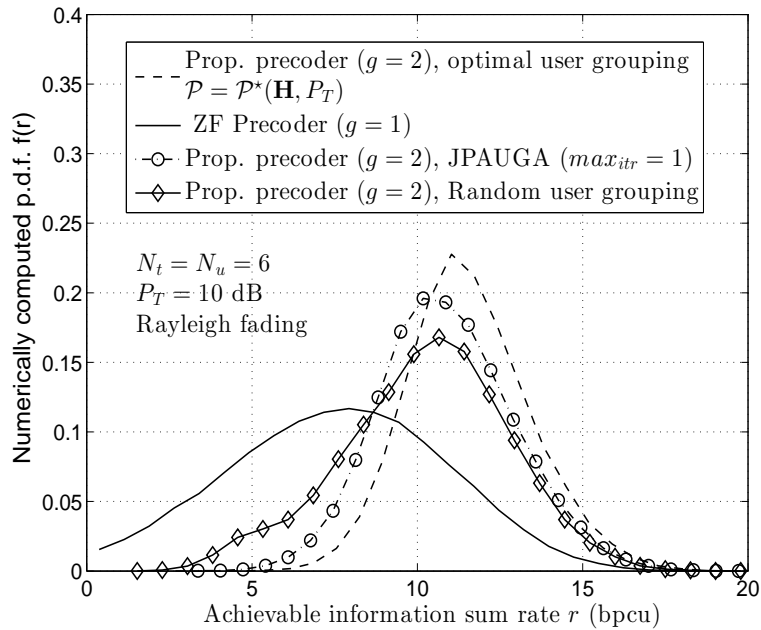, width=105mm,height=90mm}
\end{center}
\vspace{-3mm}
\caption{Numerically computed probability density function (p.d.f.) of different precoders
for a $N_t = N_u = 6$ i.i.d. Rayleigh faded channel with $P_T = 10$ dB.}
\label{fig_11}
\vspace{-5mm}
\end{figure}

\begin{figure}[t]
\begin{center}
\hspace{-1mm}
\epsfig{file=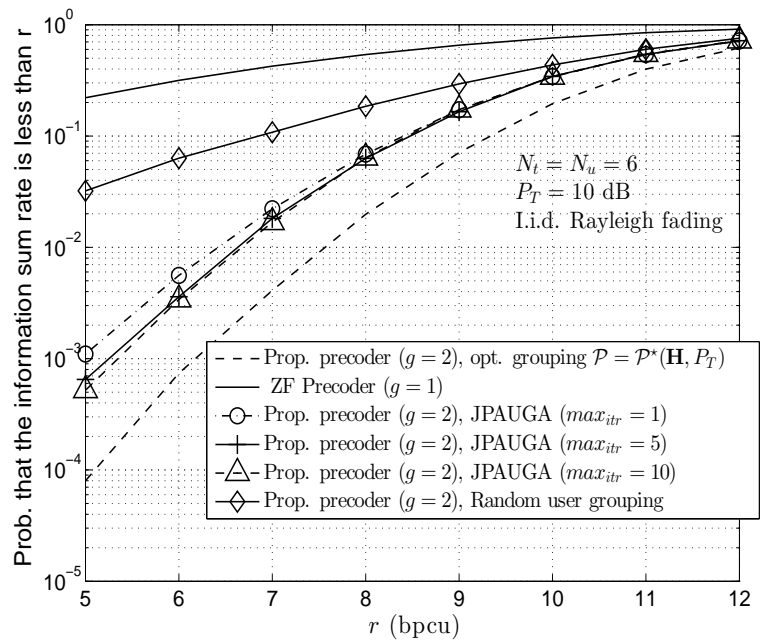, width=105mm,height=90mm}
\end{center}
\vspace{-3mm}
\caption{Probability of the event that the instantaneous sum rate is below a given sum rate $r$.
$N_t = N_u = 6$, i.i.d. Rayleigh fading with $P_T = 10$ dB.}
\label{fig_41}
\vspace{-5mm}
\end{figure}

In this section we consider an i.i.d. Rayleigh fading channel, i.e., the channel
gains $h_{k,i}^*$ are i.i.d. ${\mathcal C}{\mathcal N}(0,1)$.
In Fig.~\ref{fig_11} we consider a $N_t = N_u = 6$ i.i.d. Rayleigh fading channel with $P_T = 10$ dB, for which we numerically compute and plot the probability density function (p.d.f.)
of the sum rate achieved by the ZF precoder (i.e., $r =C_{ZF}({\bf H}, P_T)$), the proposed user grouping precoder
with optimal user pairing (i.e., $r = r^{\bigstar}({\bf H}, P_T)$ with $g=2$), the proposed precoder with random user pairing\footnote{\footnotesize{Pairs of users ($g=2$) being chosen randomly independent of the channel realization, followed by optimal waterfilling power allocation for the randomly chosen
user pairing.}}, and
the proposed precoder with JPAUGA ($g=2$ and $max_{itr} = 1$).
The achievable sum rate for each precoder is random due to the random channel gains. 
It can be observed from the figure that the probability of the sum rate assuming small values (compared to the mean value, i.e., ergodic rate) is much higher
for the ZF precoder than for the proposed user grouping based precoders.
For example, the sum rate of the ZF precoder is less than $6$ bpcu with a probability of $0.2$ (i.e., for every fifth channel realization on an average),
whereas the sum rate achieved by the proposed precoder based on JPAUGA user pairing ($\max_{itr} = 1$) falls below $6$ bpcu with a probability less than $0.01$ (i.e., one in hundred channel realizations).
Therefore, in a way the proposed user grouping based precoders improve the conditioning of the channel. 
%This is due to the fact that when the channel
%is ill-conditioned the channel matrix is almost rank deficient and therefore simple
%channel pre-inversion (i.e., ZF precoding) at the BS leads to small magnitude of the effective
%channel gains to the users.
%However by grouping users, the proposed precoder uses channel pre-inversion only to cancel inter-group interference, and this helps in significantly improving the sum rate.

We also represent the numerical data collected for Fig.~\ref{fig_11}, in terms of the
probability that a given precoding scheme achieves an instantaneous information sum rate less than some
specified rate $r$. This is shown in Fig.~\ref{fig_41}, where it can be clearly seen that
for a given fixed rate $r$, compared to the ZF precoder the proposed precoders (with $g=2$) have a significantly lower probability of the event
that the instantaneous information sum rate falls below $r$.
For any precoder let us define its critical rate $r$ to be such that the probability that its instantaneous
information sum rate falls below $r$ bpcu equals $1 \times 10^{-3}$.
It can be observed that the critical value of $r$ for the proposed precoder with JPAUGA
based user grouping (only one iteration) is $5$ bpcu which is only about $1$ bpcu less
than the critical rate of
the proposed precoder with
optimal user grouping.
Numerical simulations reveal that the critical rate of the ZF precoder
is only about $0.1$ bpcu, and therefore using the proposed precoder based on JPAUGA user grouping results in a $50$ fold increase
in the critical rate when compared to the ZF precoder.
It is noted that the proposed precoder based on JPAUGA user pairing
achieves this performance improvement at a complexity similar to the ZF precoder
(see Remark \ref{remark_same_cmplx} in Section \ref{cmplx_jpauga}).

In Fig.~\ref{fig_41}, we also plot the curves for the proposed precoder based on JPAUGA
user grouping ($g=2$), for $\max_{itr} = 5$ and $\max_{itr} = 10$.
It can be seen that the performance improves with increasing number of iterations.
However this improvement in performance is small relative to the improvement
achieved by switching from random user grouping to optimal user grouping.
This also supports the comment made in Section \ref{cmplx_jpauga}, on the fast convergence of JPAUGA.
\begin{figure}[t]
\begin{center}
\hspace{-1mm}
\epsfig{file=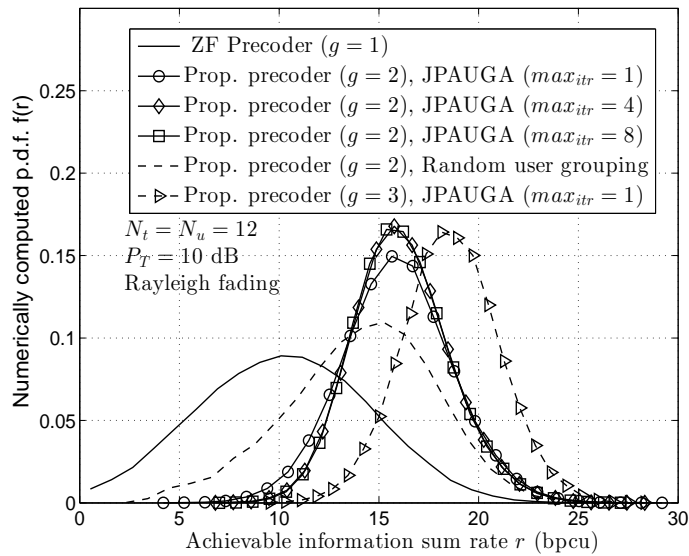, width=95mm,height=80mm}
\end{center}
\vspace{-3mm}
\caption{Numerically computed probability density function (p.d.f.) of different precoders
for a $N_t = N_u = 12$ i.i.d. Rayleigh fading channel with $P_T = 10$ dB.}
\label{fig_12}
\vspace{-5mm}
\end{figure}

In Fig.~\ref{fig_12}, we plot the numerically estimated p.d.f. of the achievable sum rate
for $N_t = N_u = 12$. We are unable to plot the p.d.f. of the sum rate achieved by the proposed precoder with
optimal user grouping due to its prohibitive complexity (with $g=2$ the number of possible groupings
is only $120$ when $N_u = 6$, but which increases to $665280$ when $N_u = 12$).
From Fig.~\ref{fig_12} we can make observations similar to that made in Fig.~\ref{fig_11}.
In Fig.~\ref{fig_12} we have also shown the p.d.f. of the proposed user grouping based on JPAUGA user grouping with $g=3$.
It is observed that by grouping $g=3$ users the p.d.f. shifts to the right when compared to $g=2$,
which implies
an even higher ergodic sum rate and an even lower probability of the sum rate being small.
This improvement in performance in going from $g=2$ to $g=3$ however comes at the cost of
increased complexity (see Section \ref{cmplx_jpauga}).
%From Fig.~\ref{fig_12} we also observe that the p.d.f.
%of the sum rate achieved by the proposed precoder is more peaky with the JPAUGA user
%grouping when compared to random user grouping. This is good in the sense that the fluctuation
%in the

\begin{figure}[t]
\begin{center}
\hspace{-1mm}
\epsfig{file=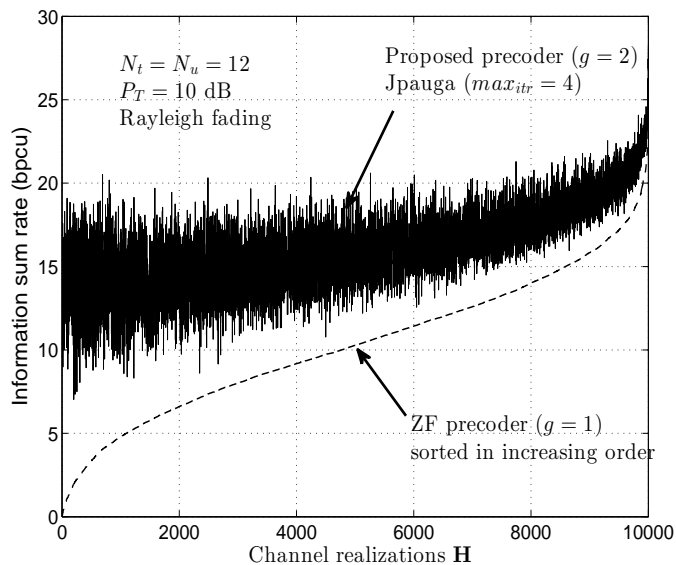, width=95mm,height=80mm}
\end{center}
\vspace{-3mm}
\caption{Sum rate of the proposed precoder with JPAUGA user grouping ($g=2$, $max_{itr} = 4$)
and the ZF precoder for ten thousand random channel realizations ($N_t = N_u = 12$, i.i.d. Rayleigh fading and $P_T = 10$ dB).
}
\label{fig_13}
\vspace{-5mm}
\end{figure}

In Fig.~\ref{fig_13} we plot the achievable sum rate of the proposed precoder (JPAUGA user grouping
with $g=2$ and $\max_{itr} = 4$) and that of the ZF precoder for ten thousand random channel realizations
($N_t = N_u = 12$, i.i.d. Rayleigh fading and $P_T = 10$ dB).
In the plot the realizations have been reordered so that the sum rate achieved by the
ZF precoder (plotted vertically) increases monotonically with the index of the ordered channel realization
(plotted horizontally). We observe that for ill-conditioned channel realizations where the
ZF precoder achieves small information sum rate, the proposed user grouping based
precoder achieves a much better performance.\footnote{\footnotesize{For channel realization
indices between $1$ and $400$ the ZF precoder achieves a sum rate less than $3$ bpcu.
For these same channel realizations the proposed user grouping based precoder
always achieves a sum rate greater than $7$ bpcu.}}

\section {Conclusions}
In this paper,
we proposed a precoding scheme in which users are grouped together in small groups of size $g$.
Multiuser beamforming is done in such a way that only {\it inter-group} interference is cancelled,
resulting in $N_u/g$ parallel non-interfering $g \times g$ Gaussian MISO broadcast channels, one
such channel for each group.
Due to the lower triangular structure of the equivalent $g \times g$ broadcast channel for each group,
successive DPC can be used to pre-cancel the intra-group interference within each group.
This method of precoding is shown to achieve a significantly better performance than the ZF
precoder, especially when the channel is ill-conditioned.
The sum rate achieved by the proposed precoder is also shown to be sensitive towards the chosen user grouping,
and therefore a novel low-complexity joint power allocation and user grouping algorithm (JPAUGA)
is proposed.

%%%%%%%%%%%%%%%%%%%%%%%%%%%%%%%%%%%%%%%%%%%%%%%%%%%%%%%%%%%%%%%%%%%%%%%%%%%%%
\appendices

\section {Proof of Theorem \ref{th1}}\label{th1_proof}
%%%%%%%%%%%%%%%%%%%%%%%%%%%%%%%%%%%%%%%%%%%%%%%%%%%%%%%%%%%%%%%%%%%%%%%%%%%%%
For a given $({\bf H}, P_T, {\mathcal P})$, from (\ref{sum_rate_tot}) it is clear that
\begin{eqnarray}
r({\bf H}, P_T, {\mathcal P}) & \geq & r({\bf H}, P_T, {\mathcal P}, {\bf p}^*)
\end{eqnarray}
since the optimal power allocation for the ZF precoder i.e., ${\bf p}^*$ (see (\ref{zf_prec_eq5})) is not necessarily the
optimal power allocation for the proposed precoder with $g \geq 2$.
Hence in order to prove (\ref{sum_cap_g2_anypair}) for any $({\bf H}, P_T, {\mathcal P})$ with the user grouping ${\mathcal P}$ having groups of size $g \geq 2$, it suffices to show that
$r({\bf H}, P_T, {\mathcal P}, {\bf p}^*)  \geq   C_{\mbox{ZF}}({\bf H}, P_T)$, i.e.
\begin{equation}
\label{sum_cap_g_anygrouping}
\sum_{k=1}^{N_u/g}  \sum_{j=1}^{g}  \log_2(1 + p_{k_j}^{*} {\mathbf R[k]}_{(j,j)}^2) \geq  C_{\mbox{ZF}}({\bf H}, P_T).
\end{equation}
Here, in the L.H.S. we have used the expression for $r({\bf H}, P_T, {\mathcal P}, {\bf p})$ from (\ref{rhpt_expr}).
In the following we will show that for any arbitrary ${\mathcal P}$ 
{
\vspace{-4mm}
\begin{equation}
\label{ug_rkjj_2}
{\mathbf R[k]}_{(j,j)}^2  \geq \frac {1} { [ ({\bf H}{\bf H}^{H})^{-1} ]_{(k_j,k_j)}  }.
\end{equation}
}
This is sufficient to prove (\ref{sum_cap_g2_anypair}) because combining (\ref{ug_rkjj_2}) and (\ref{zf_prec_eq6}), we get (\ref{sum_cap_g_anygrouping}).

%For any given grouping of users, let ${\mathcal U}_{k_j}$ be the $j$-th user in the $k$-th group.
Since ${\mathbf R[k]}$ is the upper triangular matrix in the QR-type decomposition of ${\mathbf F[k]}$, we
next examine the columns of ${\mathbf F[k]} = {\mathbf P[k]} {\mathbf G[k]}^H$.
The $j$-th column of ${\mathbf G[k]}^H$ is nothing but the complex conjugate of the channel vector of
the user ${\mathcal U}_{k_j}$. 
We firstly note that, the $j$-th column of ${\mathbf F[k]}$ is the projection of the channel vector of
user ${\mathcal U}_{k_j}$ onto ${\mathcal H}^{\perp}_k$, i.e., the space orthogonal to the space spanned by the channel vectors of users
not in the $k$-th group. Remember that for user ${\mathcal U}_{k_j}$,
${\mathcal C}_{k_j} \subset {\mathbb C}^{N_t}$ is the space of vectors orthogonal to the space spanned by the rows of ${\mathbf H[k]}$ and the rows of the previous $(j - 1)$ users in the $k$-th group (i.e., ${\mathbf h}_{k_1}^H , {\mathbf h}_{k_2}^H, \cdots , {\mathbf h}_{k_{(j-1)}}^H$). Since QR-decomposition is essentially a Gram-Schmidt orthogonalization procedure,
${\mathbf R[k]}_{(j,j)}$ is nothing but the Euclidean length of the projection of the channel vector of
user ${\mathcal U}_{k_j}$ (i.e., ${\mathbf h}_{k_j}^H$) onto the space ${\mathcal C}_{k_j}$.

In the case of ZF precoding, each group has only one user, and is therefore a special case of
the proposed user grouping scheme.  For the user ${\mathcal U}_{k_j}$, with ZF precoding,
the effective channel gain is therefore the Euclidean length of the projection of ${\mathbf h}_{k_j}^H$ onto the
space orthogonal to the space spanned by the channel vectors of the remaining $(N_u -1)$ users.
In Section \ref{prior_art}, for user ${\mathcal U}_{k_j}$, we had used ${\mathcal H}^{\perp}_{k_j}$ to denote the space orthogonal to the space spanned by the channel vectors
of the remaining $(N_u - 1)$ users.
From the definition of the space ${\mathcal C}_{k_j}$, it follows
that ${\mathcal H}^{\perp}_{k_j}$ is a subspace of ${\mathcal C}_{k_j}$.
{
\vspace{-3mm}
\begin{equation}
\label{spaces_rel}
{\mathcal H}^{\perp}_{k_j} \subset {\mathcal C}_{k_j}.
\end{equation}
}
We next show that the Euclidean length of the projection of ${\mathbf h}_{k_j}^H$ onto ${\mathcal H}^{\perp}_{k_j}$ is equal to $1/\sqrt{ [({\mathbf H} {\mathbf H}^H)^{-1}]_{k_j,k_j}}$.
Consider a row permutation matrix ${\bf T} \in {\mathbb C}^{N_u \times N_u}$, which swaps the
$k_j$-th row with the first row of any matrix with $N_u$ rows. Then the
matrix ${\bf T}{\bf H} \in {\mathbb C}^{N_u \times N_t}$ has the following structure
\begin{equation}
\label{akmat}
{\bf T}{\bf H} = \left[\begin{array}{c}
{\bf h}_{k_j}^H  \\
{\Tilde {\bf H}}
\end{array} \right]
\end{equation}
where ${\Tilde {\bf H}} = ( {\bf h}_2, {\bf h}_3, \cdots, {\bf h}_{k_{(j - 1)}}, {\bf h}_1, {\bf h}_{k_{(j+1)}}, \cdots {\bf h}_{N_u})^H$ is a sub-matrix of ${\bf H}$ containing all the rows of ${\bf H}$ except ${\bf h}_{k_j}^H$, and with ${\bf h}_1^H$ replacing ${\bf h}_{k_j}^H$ in the $k_j$-th row.
Here we also note that, ${\mathcal H}^{\perp}_{k_j}$ is the space of vectors orthogonal to the rows of ${\Tilde {\bf H}}$.
The Euclidean length of the projection of ${\bf h}_{k_j}^H$ onto the space ${\mathcal H}^{\perp}_{k_j}$
is given by
{
\vspace{-4mm}
\begin{equation}
\label{proj_zf}
c_{k_j}  = \Vert ({\bf I}_{N_t} - {\Tilde {\bf H}}^H ({\Tilde {\bf H}} {\Tilde {\bf H}}^H  )^{-1} {\Tilde {\bf H}}) {\bf h}_{k_j} \Vert 
 =  \sqrt { {\bf h}_{k_j}^H {\bf h}_{k_j} -  {\bf h}_{k_j}^H {\Tilde {\bf H}}^H ({\Tilde {\bf H}} {\Tilde {\bf H}}^H  )^{-1} {\Tilde {\bf H}} {\bf h}_{k_j}  }.
\end{equation}
}
We now consider the matrix ${\bf T}{\bf H}{\bf H}^H{\bf T}^H \in {\mathbb C}^{N_u \times N_u}$ which has the following structure.
{
\vspace{-3mm}
\small
\begin{equation}
\label{akmat_2}
{\bf T}{\bf H}{\bf H}^H{\bf T}^H = \left[\begin{array}{cc}
{\bf h}_{k_j}^H {\bf h}_{k_j}  & {\bf h}_{k_j}^H {\Tilde {\bf H}}^H\\
{\Tilde {\bf H}} {\bf h}_{k_j} &  {\Tilde {\bf H}} {\Tilde {\bf H}}^H
\end{array} \right]
\end{equation}
}
The inverse of the block partitioned matrix in (\ref{akmat_2}) is given by
{\small
\begin{equation}
\label{akmat_3}
({\bf T}{\bf H}{\bf H}^H{\bf T}^H)^{-1} = \left[\begin{array}{cc}
( {\bf h}_{k_j}^H {\bf h}_{k_j} -  {\bf h}_{k_j}^H {\Tilde {\bf H}}^H ({\Tilde {\bf H}} {\Tilde {\bf H}}^H  )^{-1} {\Tilde {\bf H}} {\bf h}_{k_j}  )^{-1}  & {\bf Y}\\
{\bf Z}  &  {\bf W}
\end{array} \right]
\end{equation}
}
with appropriate block matrices ${\bf Y}$, ${\bf Z}$ and ${\bf W}$.
Here we have used the result that for any square full rank block partitioned
matrix ${\bf V}$, of the form
{
\vspace{-3mm}
\small
\begin{equation}
\label{akmat_4}
{\bf V} = \left[\begin{array}{cc}
{\bf A} & {\bf B}  \\
{\bf C} & {\bf D}
\end{array} \right],
\end{equation}
}
the inverse is given by \cite{HornJohnson}
{
\vspace{-3mm}
\small
\begin{equation}
\label{akmat_5}
{\bf V}^{-1} = \left[\begin{array}{cc}
({\bf A} - {\bf B}{\bf D}^{-1}{\bf C})^{-1} &   -{\bf A}^{-1}{\bf B} ({\bf D} - {\bf C}{\bf A}^{-1}{\bf B})^{-1}  \\
-{\bf D}^{-1}{\bf C} ({\bf A} - {\bf B}{\bf D}^{-1}{\bf C})^{-1}  &  ({\bf D} - {\bf C}{\bf A}^{-1}{\bf B})^{-1}
\end{array} \right].
\end{equation}
}
From (\ref{proj_zf}) and (\ref{akmat_3}) it is clear that the squared Euclidean length of the projection of ${\bf h}_{k_j}^H$ onto the space orthogonal to the rows of ${\Tilde {\bf H}}$ is simply the inverse of the $(1,1)$ entry of the matrix $({\bf T}{\bf H}{\bf H}^H{\bf T}^H)^{-1}$, i.e.
{
\vspace{-4mm}
\begin{equation}
\label{akmat_6}
c_{k_j}^2 = \frac {1} { [ ({\bf T}{\bf H}{\bf H}^H{\bf T}^H)^{-1} ]_{(1,1)}}.
\end{equation}
}
Since ${\bf T}$ swaps the $k_j$-th and the first row of ${\bf H}$, it follows that
\begin{eqnarray}
\label{akmat_7}
[ ({\bf T}{\bf H}{\bf H}^H{\bf T}^H)^{-1} ]_{(1,1)}&  = & [ {\bf T} ({\bf H}{\bf H}^H)^{-1} {\bf T}^H ]_{(1,1)} 
 =  [ ({\bf H}{\bf H}^H)^{-1}  ]_{(k_j,k_j)}.
\end{eqnarray}
Combining (\ref{akmat_6}) and (\ref{akmat_7}), we have
{
\vspace{-6mm}
\begin{equation}
\label{akmat_8}
c_{k_j} = \frac {1} {\sqrt{[ ({\bf H}{\bf H}^H)^{-1}  ]_{(k_j,k_j)}} }.
\end{equation}
}
For the proposed user grouping algorithm, for any arbitrary grouping,
the projection of the channel vector of user ${\mathcal U}_{k_j}$ (i.e., ${\mathbf h}_{k_j}^H$ ) onto
the subspace ${\mathcal C}_{k_j}$ is equal to ${\mathbf R[k]}_{(j,j)}$.
From (\ref{akmat_8}), the projection of ${\bf h}_{k_j}^H$ onto the subspace ${\mathcal H}^{\perp}_{k_j}$
is equal to $1/{\sqrt{[ ({\bf H}{\bf H}^H)^{-1}  ]_{(k_j,k_j)}}}$.
From (\ref{spaces_rel}), it follows that ${\mathcal H}^{\perp}_{k_j}$ is a subspace of ${\mathcal C}_{k_j}$,
which implies that the projection of ${\bf h}_{k_j}^H$ onto ${\mathcal H}^{\perp}_{k_j}$ has a smaller Euclidean
length than its projection on ${\mathcal C}_{k_j}$\footnote {\footnotesize {The fact used here is that, the Euclidean length of the projection of any vector onto a subspace ${\mathcal B} \subset {\mathcal G}$ is smaller than its projection onto the original space ${\mathcal G}$. This can be proved using elementary linear algebra.}}.
From the above arguments,
{
\vspace{-3mm}
\begin{equation}
\label{ug_rkjj_3}
{\mathbf R[k]}_{(j,j)}  \geq \frac {1} { \sqrt {[ ({\bf H}{\bf H}^{H})^{-1} ]_{(k_j,k_j)} } }
\end{equation}
}
which proves (\ref{ug_rkjj_2}) and subsequently (\ref{sum_cap_g_anygrouping}).
$\hfill\blacksquare$
%----------------------------------------------------------

\section{Proof of Lemma \ref{lemma_1}}\label{appen_th_analysis}

Towards proving Lemma \ref{lemma_1}, we firstly observe that the ZF precoder is a special case of the proposed precoder with $g=1$.
Further it is trivial to show that for the proposed precoder with $g=2$, out of the two users in any given pair, one user (to be precise, user ${\mathcal U}_{k_2}$ for the $k$-th pair) has {\it exactly} the same channel gain as it would have had
if ZF precoding were to be used. The ``other'' user in the pair  (i.e., user ${\mathcal U}_{k_1}$ for the $k$-th pair) has a larger effective channel gain magnitude compared to its effective channel gain
if the ZF precoder were to be used.\footnote{\footnotesize{This follows from the proof of Theorem \ref{th1}.}}
For notational simplicity, let the effective channel gain of the user ${\mathcal U}_{k_1}$ be denoted by $a_k({\bf H})$ when precoding
with $g=2$ (i.e, the proposed precoder with users grouped in pairs) and by $b_k({\bf H})$ when precoding with the ZF precoder (i.e., $g=1$).
%It will be shown later than $a_k({\bf H}) \geq b_k({\bf H})$ for any realization of ${\bf H}$.
We are interested in evaluating the difference in the ergodic sum rates achieved by the proposed precoder when precoding with
$g=2$ and with $g=1$ respectively.
Since user ${\mathcal U}_{k_2}$ of the $k$-th pair has the same rate irrespective of whether $g=1$ or $g=2$, 
the difference in the ergodic sum rates is given by
\begin{equation}
\label{diff_erg_sum_rate}
d(P_T,N_u) = \sum_{k=1}^{N_u/2} {\Bigg (} {\mathbb E} {\Big [} \log_2 {\big (} 1 + \frac{P_T}{N_u}a_k({\bf H})^2{\big )}  {\Big ]}  -   {\mathbb E} {\Big [} \log_2 {\big (}1 + \frac{P_T}{N_u}b_k({\bf H})^2 {\big )}  {\Big ]} {\Bigg )}.
\end{equation}
The expectation in (\ref{diff_erg_sum_rate}) is over the distribution of ${\bf H}$.
Further, due to i.i.d. fading statistics and the fact that the pairing of users is independent of the channel realization, it turns out that
the $N_u/2$ random variables $a_k({\bf H})\,,\, k=1,2,\cdots,N_u/2$ are identically distributed, and a similar thing is true for $b_k({\bf H})\,,\, k=1,2,\cdots,N_u/2$.
Therefore, (\ref{diff_erg_sum_rate}) can be written as
\begin{equation}
\label{diff_erg_sum_rate1}
d(P_T,N_u) = \frac{N_u}{2}  {\Bigg (} {\mathbb E} {\Big [} \log_2 {\big (} 1 + \frac{P_T}{N_u}a_k({\bf H})^2 {\big )}  {\Big ]}  -   {\mathbb E} {\Big [} \log_2 {\big (} 1 + \frac{P_T}{N_u}b_k({\bf H})^2 {\big )}  {\Big ]} {\Bigg )}.
\end{equation}
With i.i.d. Rayleigh fading, twice the squared Euclidean length
of the projection of the channel vector of a given user onto the space orthogonal to the range space spanned by the channel vectors
of $N_u - g$ out of the remaining $N_u - 1$ users is $\chi^2$ distributed with $2(N_t - N_u + g)$ degrees of freedom.
This result follows immediately from the distribution of the diagonal elements of the upper triangular matrix in the QR factorization
of the i.i.d. Gaussian matrix ${\bf H}^H$ \cite{RMT}.
Further, $a_k({\bf H})$ and $b_k({\bf H})$ are nothing but the Euclidean length of the projection of ${\bf h}_{k_1}^H$ onto
the subspaces ${\mathcal C}_{k_1}$ and ${\mathcal H}_{k_1}^{\perp}$ respectively.
It can therefore be concluded that with $N_t = N_u$, $2 a_k({\bf H})^2$ and $2 b_k({\bf H})^2$ are $\chi^2$ distributed with $4$ and $2$ degrees of freedom
respectively.
Therefore, (\ref{diff_erg_sum_rate1}) can be simplified to
\begin{equation}
\label{diff_erg_sum_rate2}
d(P_T,N_u) = \frac{N_u}{2} \log_2(e) \int_0^{\infty} (x - 1) \, e^{-x} \, \log{\big (}1 + \frac{P_T}{N_u}x{\big )} \, dx.
\end{equation}
After some algebraic manipulations, we have
\begin{equation}
\label{diff_erg_sum_rate3}
d(P_T,N_u) = \frac{N_u}{2} \log_2(e) {\Big (} 1 - \frac{N_u}{P_T}\, {e^{\frac{N_u}{P_T}}} \, E_1{\big (}\frac{N_u}{P_T}{\big )} {\Big )}
\end{equation}
where $E_1(z) \Define \int_z^{\infty} e^{-t}/t \, dt$ is the exponential integral.
For $z > 0$ it is known that \cite{abram}
\begin{equation}
\label{diff_erg_sum_rate4}
\frac{1}{2} \log(1 + \frac{2}{z}) < e^{z} E_1(z) < \log(1 + \frac{1}{z}).
\end{equation}
Using (\ref{diff_erg_sum_rate4}) in (\ref{diff_erg_sum_rate3}) with $z = N_u / P_T$, we have 
\begin{equation}
\label{diff_erg_sum_rate5}
\frac{N_u}{2} \log_2(e) {\Big (} 1 - \frac{N_u}{P_T} \log{\big (}1 + \frac{P_T}{N_u}{\big )} {\Big )} \,  < \,  d(P_T,N_u) \,  <  \, \frac{N_u}{2} \log_2(e) {\Big (} 1 - \frac{N_u}{2P_T} \log(1 + \frac{2P_T}{N_u}) {\Big )}
\end{equation}
which proves the theorem. $\hfill\blacksquare$

\section {A Numerical illustration of GUGA}\label{app_num_ex}

For the sake of clarity, we now go through the steps of the proposed GUGA algorithm for the
ill-conditioned channel matrix given by
(\ref{illcond_eq}).
The transmit SNR is fixed to $P_T=29$ dB, and let the group size be fixed to $g=2$.
Further, let the given power allocation ${\bf p}$ be the ZF power allocation i.e.
\begin{equation}
\label{zf_palloc_ex}
{\bf p} \, = \, {\bf p}^* = ( 57.13 \,,\, 246.95 \,,\, 245.29 \,,\, 0 \,,\, 244.96 \,,\, 0).
\end{equation}
The first step of GUGA is to enumerate the rate of all possible ordered groups of $g$ users. For the specific case of $g=2$,
a group is essentially an ordered pair of users, and therefore the rate of all possible pairs of users can be pictorially depicted using a
$N_u \times N_u$ rate matrix whose $(i,j)$-th entry is the rate ${\mathcal I}(\{ {\mathcal U}_i, {\mathcal U}_j\})$ of the ordered pair $\{ {\mathcal U}_i, {\mathcal U}_j\}$.
We shall now go through the computation of one such ordered pair $\{{\mathcal U}_1, {\mathcal U}_5\}$.
Without loss of generality, let us assume $\{{\mathcal U}_1, {\mathcal U}_5\}$ to be the $i$-th ordered pair in some grouping.
From (\ref{weight_func_g}) it is clear that, for evaluating ${\mathcal I}(\{{\mathcal U}_1, {\mathcal U}_5\})$  we need to first compute ${\mathbf R[i]}$.
For the ordered pair $\{{\mathcal U}_1, {\mathcal U}_5\}$, ${\mathbf H[i]}$ and ${\mathbf G[i]}$ are given by
{\footnotesize
\begin{equation}
\label{Hi_Gi_1_5}
{\mathbf H[i]} = 
\left [ 
\begin{array}{cccccc}
0  & \frac{1}{2}   &  -\frac{1}{\sqrt{2}}  & \frac{1}{2}  & 0 & 0 \\
0 & -\frac{1}{2}  & 0 & 0 & -\frac{1}{\sqrt{2}} &  \frac{1}{2} \\
-\frac{1}{2} & 0 & 0 & \frac{1}{\sqrt{2}} & 0 &  -\frac{1}{2} \\
0 & 0 & 0 & -\frac{1}{\sqrt{2}} & -\frac{1}{{2}} & \frac{1}{{2}}
\end{array} \right ] \,\,,\,\,
{\mathbf G[i]} = 
\left [ 
\begin{array}{cccccc}
\frac{1}{2} & 0 & 0 & -\frac{1}{2} & \frac{1}{\sqrt{2}} & 0 \\
\frac{1}{2} & 0 & \frac{1}{\sqrt{2}}  & 0 & \frac{1}{{2}} & 0 \\
\end{array} \right ].
\end{equation}
}
${\mathbf P[i]}$ is then given by (\ref{Pi_def}).
Since ${\mathbf F[i]} = {\mathbf P[i]} {\mathbf G[i]}^H = {\mathbf Q[i]} {\mathbf R[i]} $,
we can derive ${\mathbf R[i]} $ from the Cholesky decomposition of the $2 \times 2$ matrix ${\mathbf F[i]}^H {\mathbf F[i]}$.
%Note that, since $g=2$, ${\mathbf F[i]}^H {\mathbf F[i]}$ is only a $2 \times 2$ matrix, whose
%Cholesky decomposition is trivial. Once ${\mathbf R[i]}$ is known, ${\mathbf Q[i]}$ is simply equal to
%${\mathbf F[i]} {\mathbf R[i]} ^{-1}$.
After all necessary calculations, ${\mathbf R[i]}$ is given by
\begin{equation}
{\mathbf R[i]} = 
\left [ 
\begin{array}{cc}
0.218 & -0.432 \\
0     &  0.133  
\end{array} \right ].
\end{equation}
From (\ref{weight_func_g}) it then follows that
\begin{equation}
{\mathcal I}(\{{\mathcal U}_1, {\mathcal U}_5\}) = \log_2(1 + R[i]_{(1,1)}^2 p^*_1) + \log_2(1 + R[i]_{(2,2)}^2 p^*_5) = 4.31 \,\, \mbox{bpcu}. 
\end{equation}
The rate of all possible ordered pair of users can be calculated in a similar manner.
The matrix containing the rates of all the possible ordered pairs is then given by
{\footnotesize
\begin{equation}
\label{weight_mat_ex}
{\bf I}^{(0)} = 
\left [ 
\begin{array}{cccccc}
\xout{-1} & 4.9 & 5.4 & 4.5 & 4.3 & 3.2 \\
6.7  & \xout{-1}  &  8.4  & 6.8  & 9.4 & 7.0 \\
7.3  & 8.4  & \xout{-1} & 6.4 & 7.8 &  5.8 \\
0.3 & 2.4 & 2.4 & \xout{-1} & 2.4 &  0 \\
6.0 & 9.4 & 7.8  & 6.4 & \xout{-1} & 6.7 \\
0.3 & 2.4 & 2.4 & 0 & 2.4 & \xout{-1}
\end{array} \right ].
\end{equation}
}
We note that in general the rate matrix is not symmetric, since the rate of a pair
is dependent on the ordering of the two users in that pair.
Since, the two users in a pair must be distinct the diagonal entries of the matrix in (\ref{weight_mat_ex}) are not meaningful
and are therefore crossed out.
Also, the numerical values in (\ref{weight_mat_ex}) has been rounded off to one decimal place.
Starting with the $k=0$-th iteration, ${\mathbb V}^{(0)} = {\mathcal S}$ and ${\mathbb E}^{(0)}$ is the set of all possible ordered pairs of users (${\mathbb V}^{(k)}$ and ${\mathbb E}^{(k)}$ are defined in Section \ref{guga_ref}).
The rate of the ordered pair $\{{\mathcal U}_i, {\mathcal U}_j\}$ is ${\bf I}^{(0)}_{(i,j)}$.

In the first iteration of GUGA, we search for the entry of ${\bf I}^{(0)}$ having maximum value.
From (\ref{weight_mat_ex}), it is clear that the maximum rate is that of the $(2,5)$-th entry, and hence
the first pair of users is (see (\ref{max_edge_g}))
\begin{equation}
\label{prp_s1}
{\Tilde {\mathcal S}_1} = \{{\mathcal U}_2, {\mathcal U}_5\}.
\end{equation}
Since, the second and the fifth user have already been paired, they must be removed from the active list of users, since
in any grouping each user must be paired exactly once.
The modified active set of users after the first iteration is given by
\begin{equation}
{\mathbb V}^{(1)} = \{{\mathcal U}_1\,,\, {\mathcal U}_3 \,,\, {\mathcal U}_4 \,,\, {\mathcal U}_6\}.
\end{equation}
Since the second and the fifth users are no more active, a pair which contains any one of them, cannot be chosen to be
the next pair. Therefore the next pair can only be one among the following set of active pairs
{\footnotesize
\begin{eqnarray}
\label{active_pairs_1}
{\mathbb E}^{(1)} & = & {\Bigg \{} \{ {\mathcal U}_1, {\mathcal U}_3 \} \,,\, \{ {\mathcal U}_3, {\mathcal U}_1 \} \,,\, \{ {\mathcal U}_1, {\mathcal U}_4 \} \,,\, \{ {\mathcal U}_4, {\mathcal U}_1 \} \,,\,  \{ {\mathcal U}_1, {\mathcal U}_6 \} \,,\, \{ {\mathcal U}_6, {\mathcal U}_1 \} \,,\, \nonumber \\
& & \{ {\mathcal U}_3, {\mathcal U}_4 \} \,,\, \{ {\mathcal U}_4, {\mathcal U}_3 \} \,,\,
\{ {\mathcal U}_3, {\mathcal U}_6 \} \,,\, \{ {\mathcal U}_6, {\mathcal U}_3 \} \,,\, \{ {\mathcal U}_4, {\mathcal U}_6 \} \,,\, \{ {\mathcal U}_6, {\mathcal U}_4 \} {\Bigg \}}.
\end{eqnarray}
}
A nice way to visualize this is by crossing out the second and fifth rows and columns of the weight matrix ${\bf I}^{(0)}$.
The new rate matrix is given by
{\footnotesize
\begin{equation}
\label{weight_mat_ex2}
{\bf I}^{(1)} = 
\left [ 
\begin{array}{cccccc}
\xout{-1} & \xout{4.9} & 5.4 & 4.5 & \xout{4.3} & 3.2 \\
\xout{6.7}  & \xout{-1}  &   \xout{8.4}  & \xout{6.8}  & \xout{9.4} & \xout{7.0} \\
7.3  & \xout{8.4}  & \xout{-1} & 6.4 & \xout{7.8} &  5.8 \\
0.3 & \xout{2.4} & 2.4 & \xout{-1} & \xout{2.4} &  0 \\
\xout{6.0} & \xout{9.4} & \xout{7.8}  & \xout{6.4} & \xout{-1} & \xout{6.7} \\
0.3 & \xout{2.4} & 2.4 & 0 & \xout{2.4} & \xout{-1}
\end{array} \right ].
\end{equation}
}
For choosing the next pair of the proposed pairing, we need to find the {\it non-crossed out} entry of
${\bf I}^{(1)}$ having maximum rate.
From (\ref{weight_mat_ex2}) the maximum weight non-crossed out entry is $(3,1)$ and therefore the next pair
in the proposed grouping is
\begin{equation}
\label{prp_s2}
{\Tilde {\mathcal S}_2} = \{{\mathcal U}_3, {\mathcal U}_1\}.
\end{equation}
Going ahead in a similar manner, it can be shown that the last pair is
\begin{equation}
\label{prp_s3}
{\Tilde {\mathcal S}_3} = \{{\mathcal U}_4, {\mathcal U}_6\}.
\end{equation}
Therefore, combining (\ref{prp_s1}),(\ref{prp_s2}) and (\ref{prp_s3}), the grouping proposed by the GUGA algorithm is given by
\begin{equation}
\label{prp_P}
{\Tilde {\mathcal P}} = {\Bigg \{} {\Big \{}  {\mathcal U}_2, {\mathcal U}_5 {\Big \}} \,,\, {\Big \{}  {\mathcal U}_3, {\mathcal U}_1 {\Big \}} \,,\,   {\Big \{}  {\mathcal U}_4, {\mathcal U}_6 {\Big \}} {\Bigg \}}.
\end{equation}

\section {Efficient computation of the the effective channel matrix ${\mathbf R[k]}^H$ and the beamforming matrix ${\mathbf Q[k]}$ for any arbitrary ordered group ${\mathcal S}_k = \{ {\mathcal U}_{k_1}, {\mathcal U}_{k_2}, \cdots , {\mathcal U}_{k_{g}} \}$.}
\label{th3}

Since the proposed JPAUGA needs to compute the rate ${\mathcal I}(\cdot)$ for all possible groups of $g$-users, we propose an efficient method to compute ${\mathbf R[k]}$ for any arbitrary group of users.
This is discussed in Section \ref{subsec_Rk}.
Once the user grouping and power allocation is decided by JPAUGA, the group-wise
beamforming matrices ${\mathbf Q[k]} \,,\, k=1,2,\ldots,N_g$ need to be computed.
From (\ref{qrd}) we know that ${\mathbf F[k]} = {\mathbf Q[k]} {\mathbf R[k]}$, and therefore
${\mathbf Q[k]}$ can be computed from the QR decomposition of ${\mathbf F[k]}$.
Efficient computation of ${\mathbf F[k]}$ and its QR-decomposition is discussed in Section \ref{subsec_Fk}.

\subsection{Computation of ${\mathbf R[k]}$ from ${\Big (} {\bf H} {\bf H}^H {\Big )}^{-1}$}
\label{subsec_Rk}

For the ordered group of users ${\mathcal S}_k = \{ {\mathcal U}_{k_1}, {\mathcal U}_{k_2}, \cdots , {\mathcal U}_{k_{g}} \}$, consider the row permutation matrix ${\mathbf T[k]}$ such that
{\small
\begin{equation}
\label{akamat_11}
{\mathbf T[k]}{\bf H} = \left[\begin{array}{c}
{\mathbf G[k]}  \\
{\mathbf H[k]}
\end{array} \right].
\end{equation}
}
Let ${\mathbf A[k]} \in {\mathbb C}^{N_u \times g}$ denote the matrix consisting of
only the first $g$ columns of $({\mathbf T[k]}{\bf H}{\bf H}^H{\mathbf T[k]}^H)^{-1}$.  
Using the expression for the inverse of block partitioned matrices in (\ref{akmat_5}),
${\mathbf A[k]}$ is given by
{\small
\begin{equation}
\label{akamat_12}
{\mathbf A[k]} = \left[\begin{array}{c}
{\Big (}{\mathbf G[k]} {\mathbf G[k]}^H - {\mathbf G[k]} {\mathbf H[k]}^H ({\mathbf H[k]}{\mathbf H[k]}^H)^{-1} {\mathbf H[k]} {\mathbf G[k]}^H {\Big )}^{-1}   \\
-({\mathbf H[k]}{\mathbf H[k]}^H)^{-1} {\mathbf H[k]} {\mathbf G[k]}^H {\Big (} {\mathbf G[k]} {\mathbf G[k]}^H - {\mathbf G[k]} {\mathbf H[k]}^H ({\mathbf H[k]}{\mathbf H[k]}^H)^{-1} {\mathbf H[k]} {\mathbf G[k]}^H {\Big )}^{-1}
\end{array} \right].
\end{equation}
}
Next, we make an important observation that
${\mathbf F[k]}^H{\mathbf F[k]}$ is nothing but the inverse
of the upper $g \times g$ sub-matrix of ${\mathbf A[k]}$. That is
\begin{eqnarray}
\label{fkhfk}
{\mathbf F[k]}^H{\mathbf F[k]} & {(a) \atop = } & {\mathbf G[k]} {\mathbf P[k]} {\mathbf G[k]}^H \nonumber \\
& = &  {\Big (} {\mathbf G[k]} {\mathbf G[k]}^H - {\mathbf G[k]} {\mathbf H[k]}^H ({\mathbf H[k]}{\mathbf H[k]}^H)^{-1} {\mathbf H[k]} {\mathbf G[k]}^H {\Big )} \nonumber \\
& {(b) \atop = } &  \mbox{inverse of the upper $g \times g$ sub-matrix of ${\mathbf A[k]}$}
\end{eqnarray}
where step (a) follows from the fact that ${\mathbf F[k]} = {\mathbf P[k]} {\mathbf G[k]}^H$
and step (b) follows from (\ref{akamat_12}).
From (\ref{qrd}) we know that ${\mathbf F[k]}^H{\mathbf F[k]} = {\mathbf R[k]}^H {\mathbf R[k]}$ and
therefore ${\mathbf R[k]}$ can be computed from the Cholesky factorization \cite{HornJohnson} of the inverse of the upper
$g \times g$ sub-matrix of ${\mathbf A[k]}$ (see (\ref{fkhfk})).
This Cholesky factorization has a complexity of $O(g^3)$.
In the following we therefore discuss the computation of the upper $g \times g$ sub-matrix of ${\mathbf A[k]}$.

We make an important note here that, even though ${\mathbf A[k]}$ consists of the first $g$ columns
of $({\mathbf T[k]}{\bf H}{\bf H}^H{\mathbf T[k]}^H)^{-1}$, we
need not explicitly compute the inverse of the matrix ${\mathbf T[k]}{\bf H}{\bf H}^H{\mathbf T[k]}^H$.
In fact $({\mathbf T[k]}{\bf H}{\bf H}^H{\mathbf T[k]}^H)^{-1}$ turns out to be a row and column permuted version
of $({\bf H}{\bf H}^H)^{-1}$. To see this, we note that since ${\mathbf T[k]}$ are permutation matrices,
${\mathbf T[k]}^H = {\mathbf T[k]}^{-1}$ and therefore
\begin{equation}
\label{akamat_13}
{\Big (} {\mathbf T[k]}{\bf H}{\bf H}^H{\mathbf T[k]}^H {\Big )}^{-1} = {\mathbf T[k]} ({\bf H}{\bf H}^H )^{-1} {\mathbf T[k]}^H.
\end{equation}
To be precise, exactly
$g$ rows and $g$ columns of $({\bf H}{\bf H}^H )^{-1}$ are permuted, and hence the complexity of computing
${\mathbf A[k]}$ from $({\bf H}{\bf H}^H )^{-1}$ is $O(g N_u)$.
Since for computing ${\mathbf R[k]}$, we are only interested in the upper $g \times g$ sub-matrix
of ${\mathbf A[k]}$, it can be concluded that the complexity of computing ${\mathbf R[k]}$
from $({\bf H}{\bf H}^H)^{-1}$ is only $O(g^3)$ (permuting $({\bf H}{\bf H}^H)^{-1}$ to get the upper $g \times g$ sub-matrix of
${\mathbf A[k]}$ has a complexity of $O(g^2)$ and that of inverting it is $O(g^3)$).

\subsection{Computation of ${\mathbf Q[k]}$ from ${\Big (} {\bf H} {\bf H}^H {\Big )}^{-1}$}
\label{subsec_Fk}

In the following we firstly show how ${\mathbf F[k]}$ can be computed efficiently from ${\mathbf A[k]}$
(see (\ref{akamat_12})). Since ${\mathbf F[k]} = {\mathbf Q[k]}$ ${\mathbf R[k]}$,
${\mathbf Q[k]}$ can then be computed from the QR-decomposition of ${\mathbf F[k]}$.

Right multiplication of ${\mathbf A[k]}$ by the inverse of its upper $g \times g$ sub-matrix gives
{\small
\begin{equation}
\label{ak_rmult}
{\mathbf A[k]}  {\Big (} {\mathbf G[k]} {\mathbf G[k]}^H - {\mathbf G[k]} {\mathbf H[k]}^H ({\mathbf H[k]}{\mathbf H[k]}^H)^{-1} {\mathbf H[k]} {\mathbf G[k]}^H {\Big )} = \left[\begin{array}{c}
{\mathbf I}_g  \\
-({\mathbf H[k]}{\mathbf H[k]}^H)^{-1} {\mathbf H[k]} {\mathbf G[k]}^H
\end{array} \right].
\end{equation}
}
The complexity of computing the inverse of the upper $g \times g$ sub-matrix of ${\mathbf A[k]}$ is
$O(g^3)$. The complexity of the right multiplication in (\ref{ak_rmult}) is
$O(g^2 N_u)$. Further pre-multiplication with ${\bf H}^H{\mathbf T[k]}^H$ gives the desired matrix ${\mathbf F[k]}$.
{\small 
\begin{eqnarray}
\label{ak_lmult}
 {\bf H}^H{\mathbf T[k]}^H {\Bigg (} {\mathbf A[k]}  {\Big (} {\mathbf G[k]} {\mathbf G[k]}^H - {\mathbf G[k]} {\mathbf H[k]}^H ({\mathbf H[k]}{\mathbf H[k]}^H)^{-1} {\mathbf H[k]} {\mathbf G[k]}^H {\Big )} {\Bigg )} \nonumber \\
 =  [ {\mathbf G[k]}^H  {\mathbf H[k]}^H   ]  \left[\begin{array}{c}
{\mathbf I}_g  \\
-({\mathbf H[k]}{\mathbf H[k]}^H)^{-1} {\mathbf H[k]} {\mathbf G[k]}^H
\end{array} \right] \nonumber \\
 =  {\mathbf G[k]}^H - {\mathbf H[k]}^H ({\mathbf H[k]}{\mathbf H[k]}^H)^{-1} {\mathbf H[k]} {\mathbf G[k]}^H 
% =  ({\mathbf I}_{N_t} - {\mathbf H[k]}^H ({\mathbf H[k]}{\mathbf H[k]}^H)^{-1} {\mathbf H[k]}) {\mathbf G[k]}^H  \nonumber \\
 =  {\mathbf P[k]} {\mathbf G[k]}^H
 =  {\mathbf F[k]}.
\end{eqnarray}
}
The complexity of matrix multiplication on the left hand side of (\ref{ak_lmult}) is $O(g N_u N_t)$.
The complexity of computing the QR-decomposition for ${\mathbf F[k]} \in {\mathbb C}^{N_t \times g}$
is $O(g^2N_t)$.
We also know from the previous section that the complexity of computing ${\mathbf A[k]}$
from ${\Big (}{\bf H} {\bf H}^H {\Big )}^{-1}$ is $O(g N_u)$.
Summing up the discussion above, it follows that the total complexity of computing ${\mathbf Q[k]}$
from ${\Big (} {\bf H} {\bf H}^H {\Big )}^{-1}$ is $O(g^3) + O(g^2N_u) + O(g N_u N_t) + O(g^2N_t)$.

\end{document}